\documentclass[aps,prc,twocolumn,groupedaddress,showpacs,showkeys]{revtex4}
\usepackage{graphicx}
\usepackage{amsmath,amssymb,enumerate}
\usepackage{color}
\def\W#1#2#3#4#5#6{
  \left\{
  \begin{array}{ccc}
    #1  & #2  & #3 \\
    #4  & #5  & #6 \\
  \end{array}
  \right\}
}
\def\9j#1#2#3#4#5#6#7#8#9{
  \left\{
  \begin{array}{ccc}
    #1  & #2  & #3 \\
    #4  & #5  & #6 \\
    #7  & #8  & #9 \\
  \end{array}
  \right\}
}

\def\bra#1{\langle #1 \rvert}
\def\ket#1{\lvert #1 \rangle}

\def\~#1{\tilde{#1}}

\newcommand\Veff{V_{\mathrm{eff}}}
\newcommand\Veffsm{V_{\mathrm{eff}}^{\mathrm{SM}}}
\newcommand\Vlowk{V_{\mathrm{low}k}}

\newcommand\fmi{\mathrm{fm}^{-1}}
\newcommand\TSO{{}^3S_1}

\newcommand\diff{\mathrm{d}}
\newcommand\Qbox{$\hat{Q}$-box}

\begin{document}


\title{Renormalization persistency of tensor force in nuclei}
\author{Naofumi Tsunoda}
\affiliation{Department of physics, the University of Tokyo, 7-3-1 Hongo,
Bunkyo-ku, Tokyo, Japan}
\author{Takaharu Otsuka}
\affiliation{Department of physics and Center for Nuclear Study,  
the University of Tokyo, 7-3-1 Hongo, Bunkyo-ku, Tokyo, Japan\\
National Superconducting Cyclotron Laboratory,
Michigan State University, East Lansing, MI, 48824, USA}
\author{Koshiroh Tsukiyama}
\affiliation{Department of physics and Center for Nuclear Study,  
the University of Tokyo, 7-3-1 Hongo, Bunkyo-ku, Tokyo, Japan}
\author{Morten Hjorth-Jensen}
\affiliation{Department of Physics and Center of Mathematics for
Applications, University of Oslo, N-0316 Oslo, Norway\\
National Superconducting Cyclotron Laboratory,
Michigan State University, East Lansing, MI, 48824, USA}

\begin{abstract}
In this work we analyze the tensor-force component of effective
interactions appropriate for nuclear shell-model studies, with 
particular emphasis on the monopole term of the interactions.
Standard nucleon-nucleon ($NN$) interactions such as AV8' and $\chi$N$^3$LO 
are tailored to shell-model studies by employing $V_{low k}$ techniques 
to handle the short-range repulsion of the $NN$ interactions and 
by applying many-body perturbation theory to incorporate 
in-medium effects. 
We show, via numerical studies of effective interactions for the $sd$
 and the $pf$ shells, 
that the tensor-force contribution to the monopole term of the effective 
interaction is barely changed by these renormalization procedures,
resulting in almost the same monopole term as the one of the bare $NN$ interactions.
We propose to call this feature {\it Renormalization Persistency}
of the tensor force, 
as it is a remarkable property of 
the renormalization and should have 
many interesting consequences in nuclear systems.  For higher multipole 
terms, this feature is maintained to a 
somewhat smaller extent.
We present general intuitive explanations for the 
Renormalization Persistency
of the tensor force, as well as analyses of 
core-polarization terms in perturbation theory.
The central force does not exhibit a similar
 Renormalization Persistency.
\end{abstract}


\maketitle



\section{Introduction}
 The nucleon-nucleon (NN) interaction is normally modelled in terms of several components, 
such as a central force, a spin-orbit force or a tensor force. These mathematical terms accommodate 
our phenomenological knowledge of the strong interaction, which, when used in a nuclear many-body context
is subjected to different renormalization procedures.
For the nuclear many-body problem, a given renormalization procedure
 leads to the derivation of an effective interaction,  starting from a bare
 realistic NN interaction.
 The so-called bare NN interactions exhibit a strong coupling between low-momentum and
 high-momentum degrees of freedom generated from short-range details of
 the interaction. By ``bare'' we mean that
 the abovementioned strong coupling is left untouched.  This coupling is included 
 only implicitly, via various renormalization procedures, in
 the effective interactions
 used in for example shell-model studies. 

 As an example  of  bare NN interactions, the Argonne
 interactions (AV), which are defined in terms of local
 operators in coordinate space, show a strong short-range
 repulsion~\cite{PhysRevC.51.38,PhysRevLett.89.182501}. The 
 resulting strong coupling 
 between low- and high-momentum modes makes the many-body problem
 highly non-perturbative. 
 On the other hand, in shell-model calculations, 
 the employed effective interactions  
 are defined for a specific configuration
 space (a strongly reduced Hilbert space), normally called the model
 space.  Therefore, the effective interactions for the shell model
 should be renormalized to include the effects of virtual excitations to
 the configurations not included in the model space.

 Although the properties and the 
 effects of the full interaction and various renormalized interactions have been
 investigated extensively over the years, we feel that there are still 
important features of the
 nuclear interaction which deserve some special attention. 
 In particular, we show here via several numerical studies, that the tensor force component
 of the bare nuclear interaction is left almost unaffected 
by various renormalization procedures. 
 The monopole component of the tensor force, a component of great interest 
 in studies of shell evolution (see
 discussion below)
 in nuclei toward the drip lines, is left almost unchanged under various renormalizations.
 This allows us 
 thereby to extract simple
 physics interpretations from complicated many-body systems. In this work we label such a lack of 
 renormalization influence as {\it Renormalization
 Persistency},  in short just R-Persistency. The
 R-Persistency is a property exhibited by specific terms  of the
 original nuclear Hamiltonian that are
 not affected, or barely affected, by the renormalization procedure.

 On the experimental side, present and future radioactive ion-beam facilities
 have made it possible to perform
 experiments that explore nuclei far from the stability line of
 the nuclear chart. 
 Many unexpected and new phenomena have been observed in such
 experiments carried out at radioactive
 ion-beam facilities worldwide. 
 One of the most striking results is the breaking of 
 the conventional shell structures in neutron-rich
 nuclei. Such shell evolution, unexpected in the past,
 is known by now to occur mainly due to an unbalanced
 neutron to proton ratio and specific
 rbital-dependent components of the nuclear forces.
 In particular, the nuclear tensor force plays a key role here, as
 proposed by one of the
 authors~\cite{PhysRevLett.95.232502,Otsuka:2009qs}. One of the
 most useful quantities to probe the effect of tensor force is the
 so-called monopole matrix element. The monopole matrix
 element~\cite{BANSAL:1964kn,Poves:1981hw} of the
 two-body interaction between  two single-particle states labelled $j$
 and $j'$ and  total two-particle isospin $T$ is defined as 
 \begin{equation}
  V_{j,j'}^T=\frac{\sum_{J}(2J+1) \bra{jj'}
   V \ket{jj'}_{JT}}{\sum_{J} (2J+1)},\label{eq:monopole}
 \end{equation}
 for $j\ne j'$~\footnote{For the $j=j'$ case, the definition is slightly
 different.}. Here $\bra{\,\cdot\,\cdot} V \ket{\,\cdot\,\cdot}_{JT} $
 denotes the anti-symmetrized two-body matrix element coupled to total
 angular momentum $J$ and total isospin $T$. The monopole matrix element
 is crucial for shell evolution, because it affects the effective single
 particle energy linearly. For instance, if $n_n(j')$ neutrons occupy
 the single-particle state $j'$, 
 they shift
 the effective single particle energy of protons in the state $j$ 
 as follows,
 \begin{align}
  \label{eq:ESPE}
  \Delta \epsilon_p(j)=\frac{1}{2}
  \left(V_{jj'}^{T=0}+V_{jj'}^{T=1}\right)
  n_n(j'),
 \end{align}
 where $\Delta \epsilon_p(j)$ represents the change of the effective
 single particle energy of protons in the single-particle state
 $j$. When we consider the tensor-force contribution, the monopole
 matrix elements always have different signs between a pair of
 spin-orbit partners. For example, the interaction matrix elements
 $V_{j_{>}j'}$ and $V_{j_{<}j'}$ have opposite sign. Here we define
 $j_{>}$ and $j_{<}$ to represent the spin-orbit partners, that is,
 $j_{>}=l + 1/2$ and $j_{<} = l-1/2$, where $l$ stands for the orbital
 momentum of a given single-particle state. In this case, the
 tensor-force changes the  spin-orbit splitting between $j_{>}$ and
 $j_{<}$. The shell structure is also altered, in particular if we have
 a sizable number of neutrons in the single-particle state $j'$.
 
 In previous studies~\cite{PhysRevLett.95.232502},
 the tensor-force component in  effective interactions for shell-model
 calculations was, for the sake of simplicity, modelled via the exchange of $\pi$  
 and $\rho$ mesons only. To a large extent, this yields results close to 
 the tensor force in realistic NN interactions. In fact,
 this model describes rather well the experimental data 
 in several mass regions~\cite{Otsuka:2009qs}. However, it is far from
 trivial that the tensor force in effective interactions for the shell-model 
 can be considered to be given by the exchange of $\pi$ and  $\rho$ mesons only.
 
 The aim of this article is thus to investigate the R-Persistency of the nuclear tensor
 force and understand the validity of the above assumption through theoretical
 studies, based on realistic NN interactions and microscopic theories for
 deriving effective interactions, focusing on the effective interaction
 for the shell model~\footnote{A short version of the present results 
 was included in~\cite{Otsuka:2009qs}, while more substantial, deeper and wider
 discussions are given in this paper.}.
 
This work is organized as follows.
 First we briefly review the theory for constructing 
 effective interactions
 in Sec.~\ref{sec:veff_th}.
 In Sec.~\ref{sec:vlowk} and Sec.~\ref{sec:veffsm}, the R-Persistency of
 the monopole part of the tensor force from various approaches to
 the effective interactions will be discussed.
 In Sec.~\ref{sec:multipole} we present not only the monopole part but
 also the two-body matrix elements including multipole part of the
 various effective interactions and discuss their tensor force
 components.
 For the sake of completeness, we include analyses using other NN
 interactions in Sec.~\ref{sec:n3lo}. The last section contains our
 conclusions.

\section{Construction of the effective interaction for the shell model}
\label{sec:veff_th}
The aim of this section is to give a 
brief sketch of the theoretical methods we employ in our analyses of the nuclear force.
 To construct the effective interactions for the nuclear shell model, we 
 use many-body perturbation theory (MBPT). However,  as inputs to MBPT,
 we cannot use bare realistic NN interactions directly, since their
 high-momentum  components make MBPT
 non-convergent~\cite{PhysRevC.65.051301}. 
 We integrate out this high-momentum components employing a 
 renormalized interaction  defined only in the low-momentum
 space below a certain sharp cutoff $\Lambda$ and designed not to change
 two-body observables like NN scattering
 data~\cite{Bogner20031}. This recipe defines a cutoff dependent family
 of interactions, normally labelled as  $\Vlowk(\Lambda)$, which to be more specific, can be written as
 \begin{equation}
  \Vlowk (\Lambda) = P_{\Lambda}V_{\mathrm{bare}}P_{\Lambda}
   + \delta V_{ct} (\Lambda),\label{eq:general}
 \end{equation}
 where $P_{\Lambda}$ indicates a projection operator onto
 the low-momentum space below $\Lambda$. The term $\delta V_{ct} (\Lambda)$
 represents the correction term coming from the renormalization procedure. 
 In other words, $P_{\Lambda}V_{\mathrm{bare}}P_{\Lambda}$ is a
 simple projection to a low-momentum space, while $\delta V_{ct} (\Lambda)$ 
 emerges as a result of the chosen renormalization procedure. 
 By construction, $\Vlowk(\Lambda)$ approaches the original NN interaction
 in the limit $\Lambda\rightarrow\infty$.
 A complete renormalization scheme would generate higher-body forces as well, 
 such as three-body and four-body forces,
 $V_{\mathrm{3N}}$ and $V_{\mathrm{4N}}$, respectively. In this
 work we limit ourselves to two-body ($V_{\mathrm{2N}}$) interactions only. Thus the cutoff
 dependence of physical quantities can be used to assess the error made
 by omitting more complicated many-body forces. The term
 $P_{\Lambda}V_{\mathrm{bare}}P_{\Lambda}$ should contain the
 long-range part of one-pion exchange interaction as a major
 component.

\begin{figure}[htbp]
  \begin{center}
   \includegraphics[width=50mm]{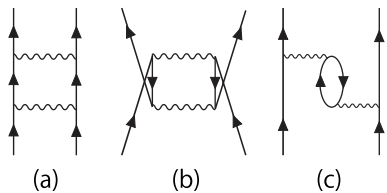}
   \caption{Examples of diagrams to second-order in the interaction
   $H_1$ included in the \Qbox.
   The diagrams are referred to as 
   (a) particle-particle ladder, (b) hole-hole ladder and 
   (c) core-polarization, respectively.}
   \label{fig:qbox_2nd}
  \end{center}
 \end{figure}

 Next, we proceed to MBPT. The low-momentum interaction $\Vlowk$ is a
 good starting point for MBPT because we can
 avoid the difficulty caused by the strong short-range repulsion. 
 For a degenerate model space, the effective interaction
 $\Veff$ can be written as
 \begin{equation}
  \Veff = \hat{Q} - \hat{Q}'\int \hat{Q}
   + \hat{Q}'\int \hat{Q}\int \hat{Q} - \cdots ,\label{eq:fd_expansion}
 \end{equation}
 where $\hat{Q}(E_0)$ is the so-called $\hat{Q}$-box, defined as
 \begin{equation}
 \hat{Q}(E_0) \equiv  PH_1 P +
  PH_1 Q \frac{1}{E_0 - QHQ}QH_1 P.
 \end{equation}
 Here, the Hamiltonian is divided into an unperturbed part $H_0$ and an
 interaction part $H_1$, 
 $H=H_0+H_1$ and the model space is set to
 be degenerate with respect to the unperturbed Hamiltonian $H_0$ with energy
 $E_0$. The integration symbols in Eq.~(\ref{eq:fd_expansion}) represent
 the inclusion to infinite order of so-called folded diagrams, see
 Refs.~\cite{Kuo_springer,HjorthJensen1995125} for details. 
 The $\hat{Q}$-box is given by diagrams which are valence-linked and
 irreducible, while $\hat{Q'}$ indicates that only diagrams which are of
 second- or higher-order in terms of the interaction $H_1$, are
 included, see for example Fig.~\ref{fig:qbox_2nd}.

 We can solve Eq.~(\ref{eq:fd_expansion}) by the following iterative formula
 \begin{equation}
  \Veff^{(n)} =  
   \hat{Q}(E_0) + \sum _{m=1}^{\infty} \hat{Q}_m(E_0) \{
   \Veff^{(n-1)}\}^{m},\label{eq:fml_itr_sol}
 \end{equation}
 where {\small $ {\displaystyle \hat{Q}_m(E_0) =
 \frac{1}{m!}(\frac{\diff^m \hat{Q}(\omega)}{\diff
 \omega^m})_{\omega=E_0} }$}. 
 In this work, we take  into account diagrams up to second or third
 order in the interaction $H_1$ for the calculation of the
 $\hat{Q}$-box. 
 
 By using this two-step method,  we can start from an arbitrary bare
 realistic NN interaction. We calculate effective interactions starting
 from AV8'~\cite{PhysRevC.51.38,PhysRevLett.89.182501} and the chiral
 $\chi$N$^3$LO interaction~\cite{RevModPhys.81.1773}. Results using the
 AV8' interaction are shown in the following sections while our  results
 obtained with the  $\chi$N$^3$LO interaction are shown in
 Sec.~\ref{sec:n3lo} for the sake of completeness.

 Finally, to extract the tensor component from the obtained effective
 interactions, we employ the spin-tensor
 decomposition employed in for example
 Refs.~\cite{Kirson1973110,PhysRevC.45.662,brown1988191}
 {\small 
 \begin{eqnarray}
  \label{eq:dec}
   \bra{abLS} V_p\ket{cdL'S'}_{J'T} =
   (-1)^{J'}\hat{p}
   \W LS{J'}{S'}{L'}p \notag \\
  \times \sum_J
   (-1)^{J}\hat{J}
   \W LSJ{S'}{L'}p
   \bra{abLS}V\ket{cdL'S'}_{JT},
 \end{eqnarray}}
 where $\bra{\,\cdot\,\cdot LS}V\ket{\,\cdot\,\cdot L'S'}_{JT}$ denotes
 the $LS$-coupled matrix element of the effective interaction. Here $a$ (as well as $bcd$)
 is shorthand for the set of quantum numbers $(n_a,l_a)$, etc. The operator
 $V_p$ is defined as the scalar product $V_p \equiv 
 U^{(p)} \cdot 
 X^{(p)}$, where $U^{(p)}$ and $X^{(p)}$ are irreducible tensors of rank 
 $p$, applying to operators in both spin and coordinate space.
 The tensor component is extracted by setting $p=2$ in
 Eq.~(\ref{eq:dec}). Finally, in the above equation 
we have defined $\hat{p}=2p+1$ and $\hat{J}=2J+1$.

 \section{Tensor force in low-momentum interaction $\Vlowk$}
 \label{sec:vlowk}
  \begin{figure*}[htb]
   \begin{center}
    \begin{tabular}{cc}
     \resizebox{80mm}{!}{\includegraphics[angle=270]
     {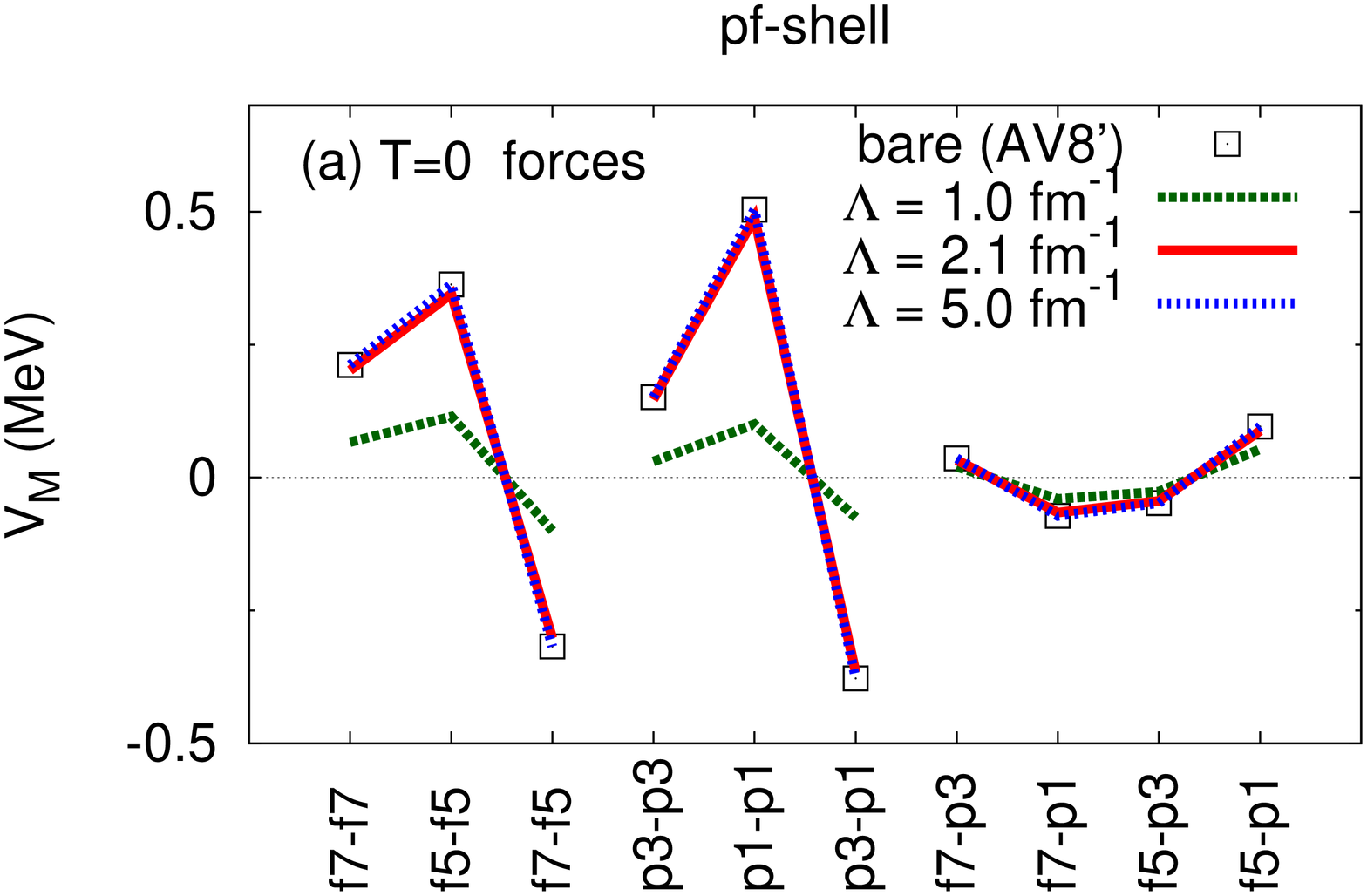}} &
     \resizebox{72.25mm}{!}{\includegraphics[angle=270]
     {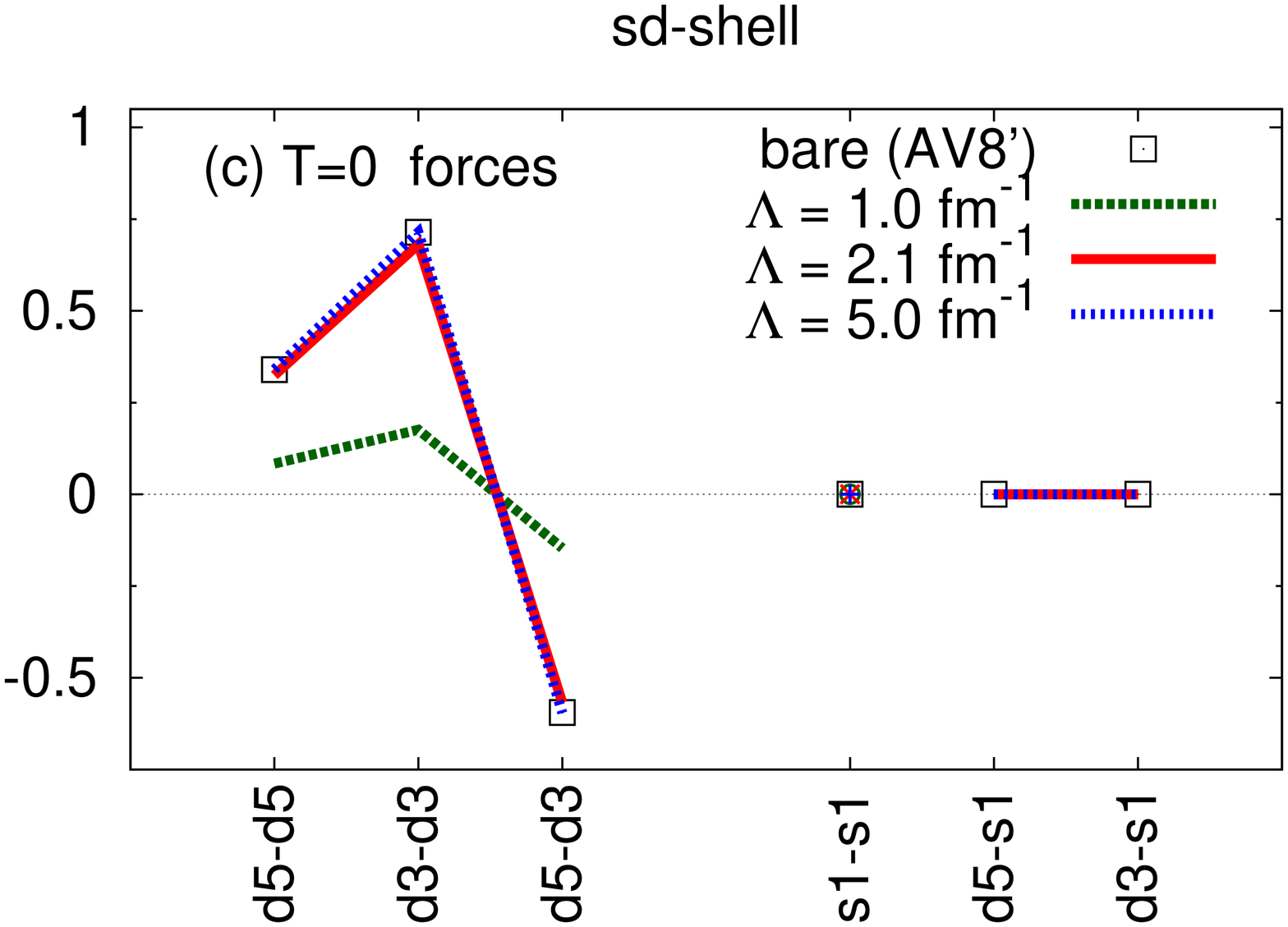}} \\
     \resizebox{80mm}{!}{\includegraphics[angle=270]
     {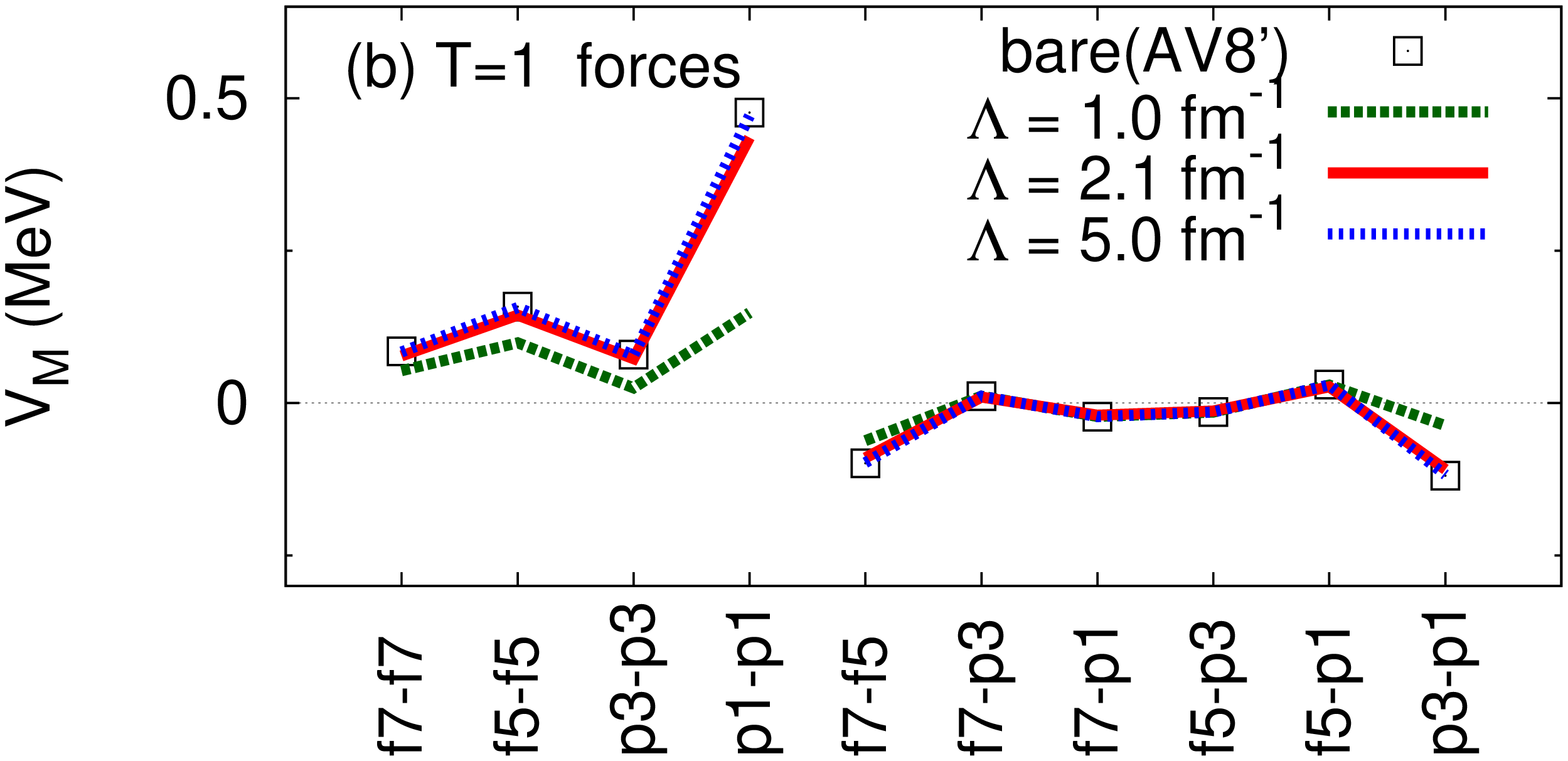}} &
     \resizebox{72.25mm}{!}{\includegraphics[angle=270]
	 {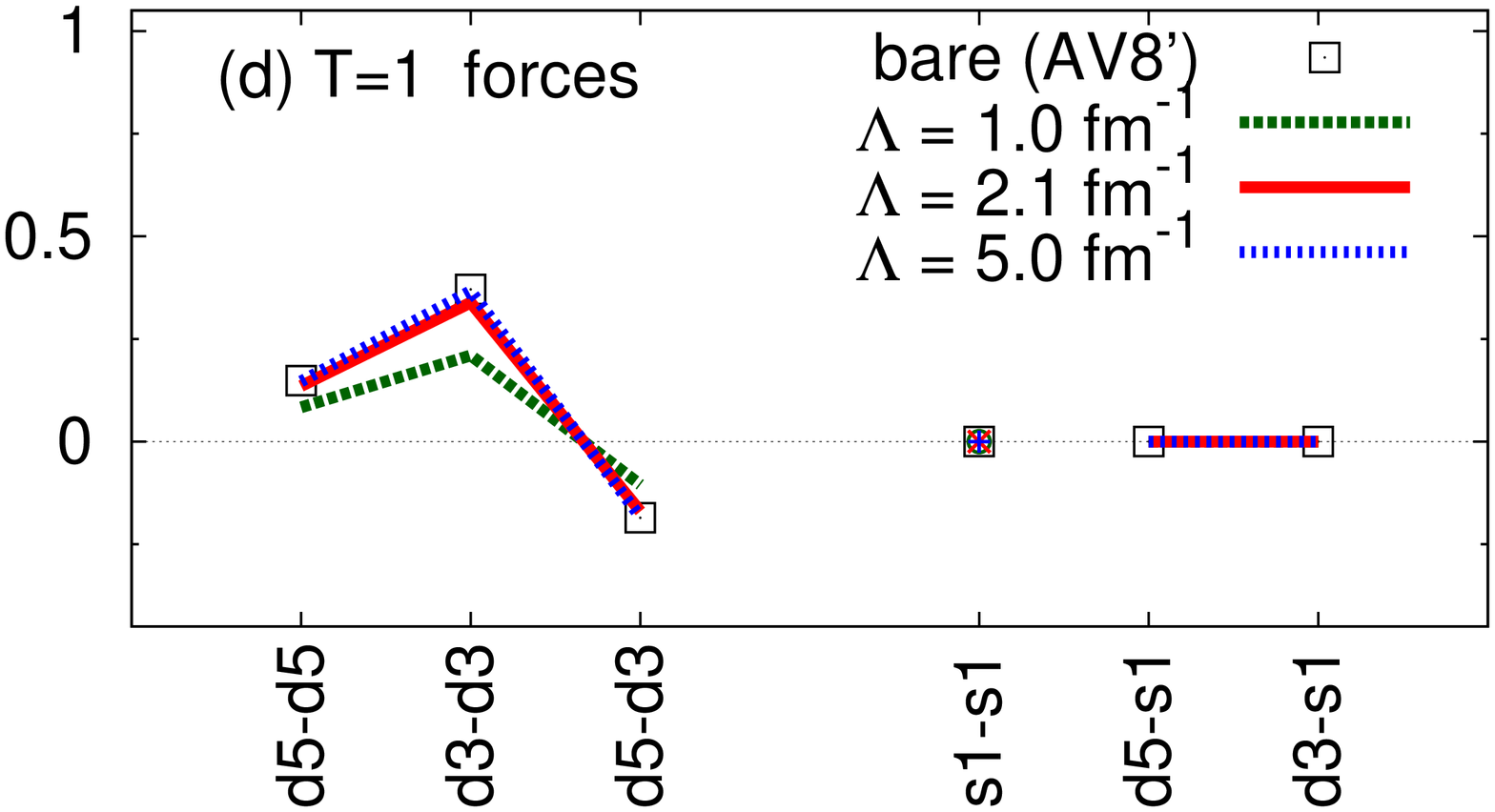}} \\
    \end{tabular}
    \caption{(color online.)
    Tensor-force
    monopole component of low-momentum interaction $\Vlowk$ as function
    of the cutoff parameter $\Lambda$ for (a) $T=0$ forces in
    $pf$-shell, (b) $T=1$ forces in $pf$-shell, (c) $T=0$ forces in
    $sd$-shell and (d) $T=1$ forces in $sd$-shell.
    The cutoff parameter $\Lambda$ of $\Vlowk$ varies from $1.0~\fmi$ to
    $5.0~\fmi$.} \label{fig:vlowk_ten}
   \end{center}
  \end{figure*}

  We now present results obtained by the theoretical methods
  described in the previous section.
  Figure~\ref{fig:vlowk_ten} shows the monopole part of the
  tensor-force of the renormalized $\Vlowk$ interaction derived 
  from the Argonne V8' (AV8') potential
  for the  $sd$-shell and the $pf$-shell. The cutoff value $\Lambda$
  varies from $1.0~\fmi$ to $5.0~\fmi$. Here, we employ units where $c =
  \hbar = \hbar^2/m = 1$. 
  The typical
  value of the cutoff is determined by the best reproduction of the
  binding energies of $^{3}\mathrm{H}$ and $^{4}\mathrm{He}$. The
  resulting cutoff value lies around
  $2.0~\fmi$~\cite{PhysRevC.70.061002}. A too small cutoff $\Lambda$
  (for example $1.0~\fmi$ in momentum space) cannot resolve the
  necessary degrees of freedom. Since the Compton length of the pion is
  approximately $0.7~\mathrm{fm}$, a cutoff $\Lambda = 1.0~\fmi$, which
  corresponds to $1.0~\mathrm{fm}$ in coordinate space, is too small to
  resolve the exchange of a pion.
  Although the resulting renormalized interaction $\Vlowk$
  with $\lambda = 1.0~\fmi$ may not contain an appropriate tensor force for
  shell-model calculations, we include its result in
  Fig.~\ref{fig:vlowk_ten} and 
  subsequent similar figures for the sake of completeness.

  We now present the results for the cutoff values   
  $\Lambda = 1.0, 2.1$ and $5.0~\fmi$ in Fig.~\ref{fig:vlowk_ten}. The
  matrix elements are calculated using a harmonic oscillator basis with
  $\hbar\omega = 14$ and $11~\mathrm{MeV}$ 
  for the $sd$-shell and  the $pf$-shell, respectively.
  Except for a very low (and thereby unreasonable) cutoff value 
  $\Lambda = 1.0~\fmi$, one finds, both in the $sd$-shell and the 
  $pf$-shell, that the monopole part of the  
  tensor force of $\Vlowk$ has almost no cutoff dependence, and
  has almost the same strength as that of the original NN interaction.
  Thus, within the usual values of the cutoff, 
  we can see that the monopole part of the tensor force 
  fulfills the R-Persistency almost perfectly with respect to the
  renormalization of the short-range part of the NN interaction.
 
  We look now into the robustness and the generality of the features 
  discussed above.
  For this purpose,
  we consider the relative motion of two interacting 
  nucleons.  The orbital angular momentum of the relative motion can 
  be $L$=0 ($S$), 1 ($P$), 2 ($D$), {\it etc}.  If the tensor force is 
  acting between two states, there is no coupling between two $S$ states, 
  because the relative coordinate operator in the tensor force is 
  of rank 2.  The $S$-to-$S$
  coupling is thus zero.  This results in strongly suppressed   
  contributions to the tensor force from the short-range part of the relative-motion 
  wave function, since a good fraction of the short-range repulsion stems from $S$ waves.
  Partial waves higher than $S$ waves carry also a centrifugal barrier component which results 
  in smaller short-range contribution to the tensor force relative to $S$ waves.
  Thus, changes of the potential at short
  distances do not affect matrix elements of the tensor force
  for low momentum states.
  This seems to be the basic reason why the tensor force remains almost
  the same throughout the renormalization procedure.  In other words,
  there is a sound reason to expect the R-Persistency for the tensor
  force regarding the treatment of the short-range correlation.
  On the other hand, the present argument may not 
  be applied to other parts of the nuclear force such as the central force.

  The second term $\delta V$ of Eq.~(\ref{eq:general}) is due to the
  renormalization. It includes for example the central-force component
  at intermediate inter-nucleon distances, and
  may affect, in principle, the tensor force as well.
  The first term, $P_{\Lambda}V_{\mathrm{bare}}P_{\Lambda}$, 
  is equal to the bare
  tensor force in the limit of $\Lambda \rightarrow \infty$ by
  definition. In this limit $\delta V$ is zero.
  Since matrix elements of the tensor force, particularly for 
  low-momentum states, are not affected much by
  the short-range modification, the effect of the tensor-force component 
  in the first term of Eq.~(\ref{eq:general}) remains the same  
  to a large extent,  even with finite $\Lambda$ values, 
  unless it becomes extremely small.
  The fact that the R-Persistency is almost  fulfilled in 
  numerical calculations (as we can see in Fig.~\ref{fig:vlowk_ten})
  implies therefore
  that the second term $\delta V$ results in small contributions to  the tensor force, 
  or does not change the long-range part of the tensor force.
  The origin of the weak tensor force component in $\delta V$ 
  can be understood by the arguments presented in Sec.~\ref{sec:veffsm}, arguments which are
based  
  on the close relation between the $\Vlowk$ renormalization 
  process and contributions from  MBPT that represent long-range corrections, as 
  discussed in~\cite{Bogner20031,Bogner:2001jn} as well.
  We shall come back to this point in Sec.~\ref{sec:veffsm}.
 
  \section{Renormalization of the central force}
  Contrary to the tensor force, it can be seen from our numerical 
  studies that the central force does not fulfill
  the R-Persistency, and is indeed affected strongly by the 
  renormalization procedure due to the short-range part of the NN interaction.
  This is reflected in a much stronger cutoff dependence as well. The
  central-force monopole part of $\delta V$ in Eq.~(\ref{eq:general}) is thus
  not small.
  \begin{figure}[htb]
   \begin{center}
    \begin{tabular}{ccc}
     \resizebox{47.62mm}{!}{\includegraphics[angle=270]{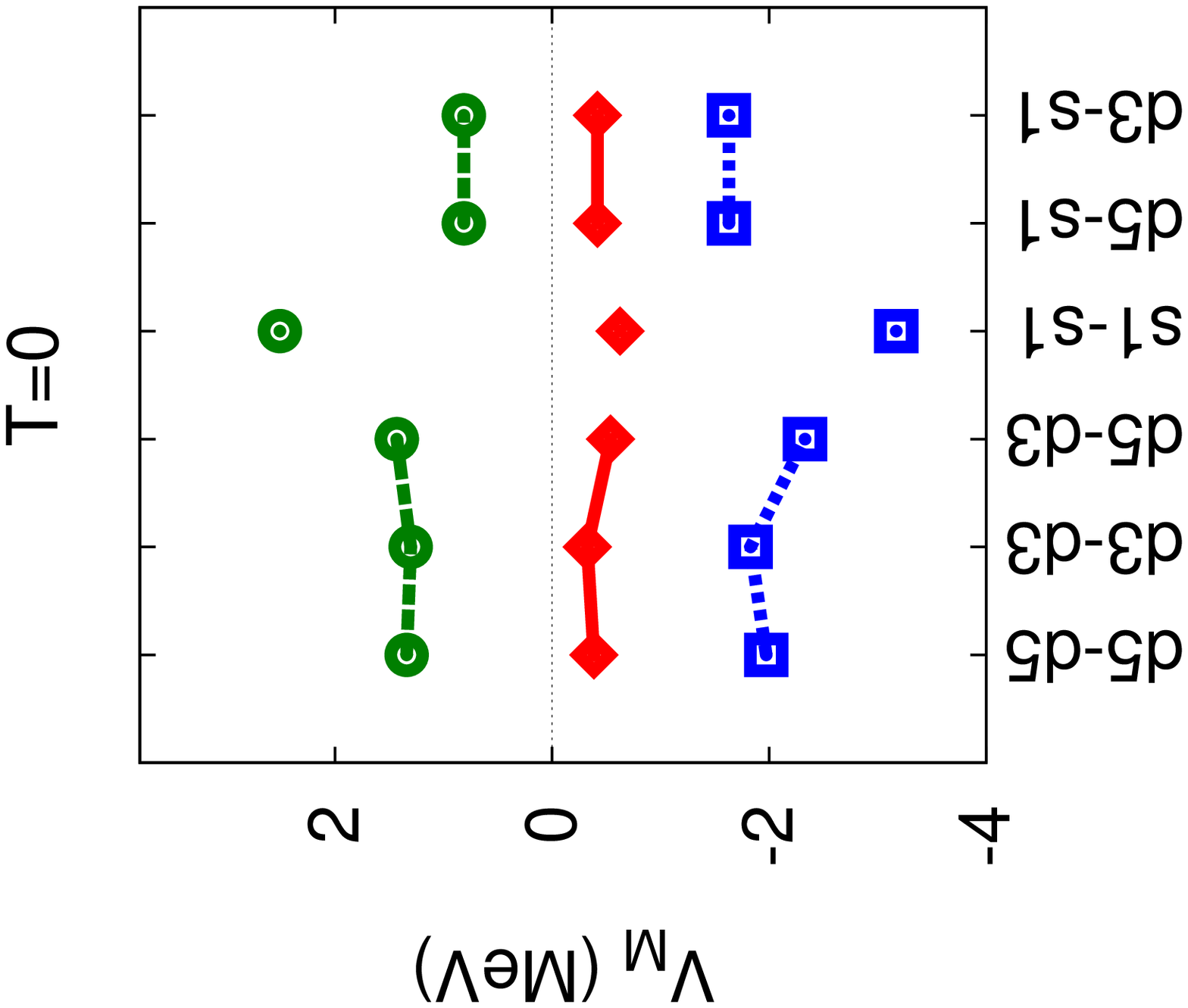}}
     & &
     \resizebox{32.38mm}{!}{\includegraphics[angle=270]{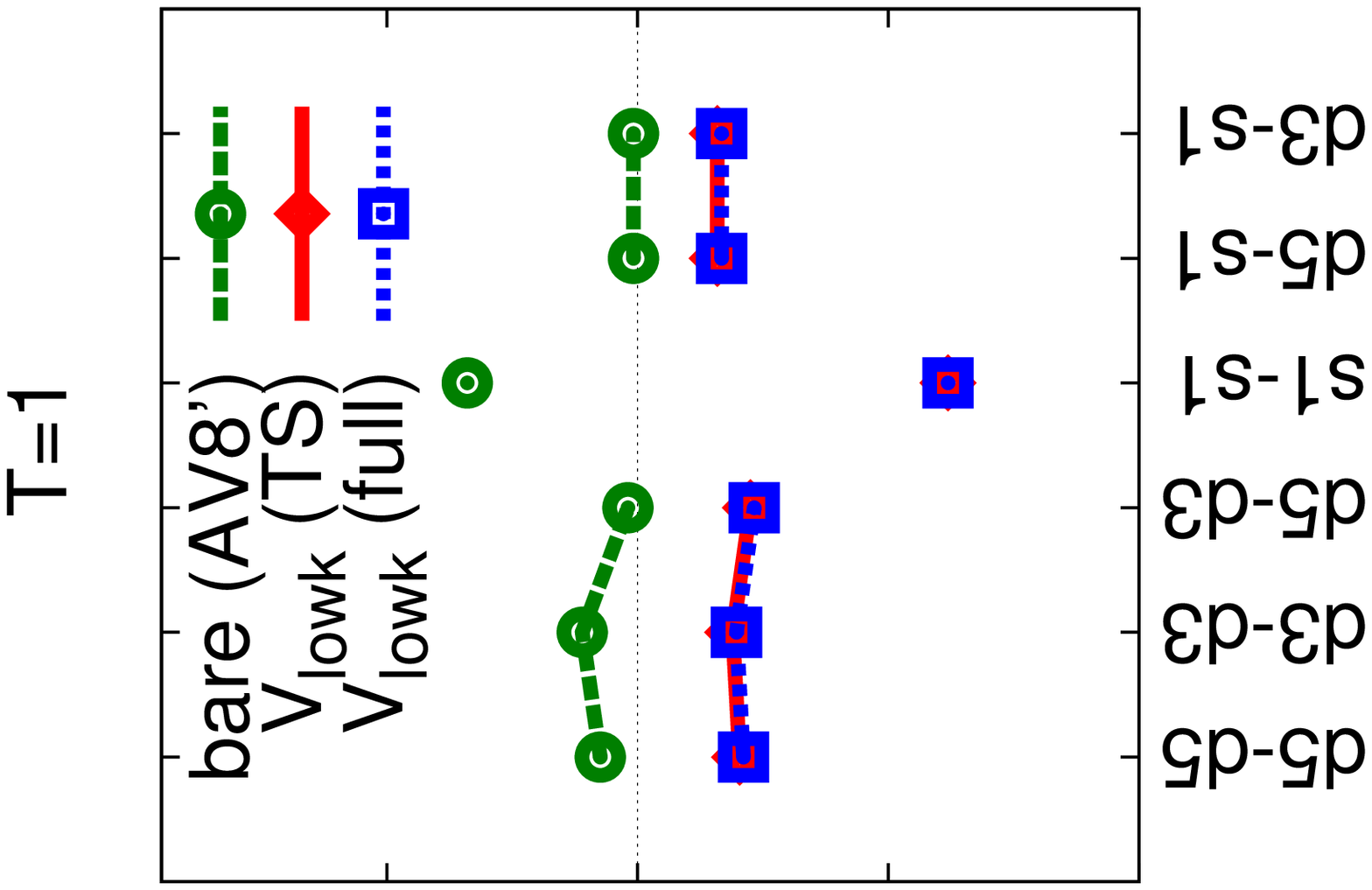}}
    \end{tabular}
    \caption{\label{fig:ren_ten} (color online.)
    Central-force component of the monopole term of the bare AV8' (bare in the figure), 
$\Vlowk$ (TS) and $\Vlowk$
    (full) for the $sd$-shell, see text for further details and discussions. 
    The central-force component is obtained using the decomposition of Eq.~(\ref{eq:dec}).
    The effect of the renormalization on the short-range tensor
    force is also shown. The cutoff value is chosen as $\Lambda = 2.1 ~\fmi$.}
   \end{center}
  \end{figure}
  
  Figure~\ref{fig:ren_ten} shows the monopole part of the central 
  force of the bare AV8' potential obtained by the decomposition of
  Eq.~(\ref{eq:dec}). In this figure we show also the corresponding
  central-force monopole component using the $\Vlowk$ renormalized
  interaction originating from the AV8' potential, labelled
  by ``full'' in the above figure.  We show also
  results where the tensor force component has been subtracted
  from the bare NN interaction in the renormalization procedure,  
  labelled by ``TS'' in the figure.
  What we can see in
  Fig.~\ref{fig:ren_ten} is the effect of the renormalization due to the
  short-range part of the bare realistic NN
  interaction. The difference between ``bare AV8' '' and ``$\Vlowk$ 
  (TS)'' lies mainly in the renormalization due to the short-range part of 
  the central force, as the tensor force is subtracted in 
  ``$\Vlowk$ (TS)''.  On the other hand, the difference between 
  ``$\Vlowk$ (TS)'' and ``$\Vlowk$ (full)'' comes solely from the
  renormalization due to the short-range part of the tensor force. 
  
  In the $T=0$ channel, the effect of the renormalization procedure on
  the short-range part of the tensor force is
  comparable to that of the central force, while in the $T=1$ channel
  this effect is almost negligible. This is a quite remarkable feature.
  Let us discuss this feature in some detail by considering the Schr\"odinger
  equation for the deuteron. The deuteron has isospin $T=0$, spin $S=1$,
  orbital momentum $L=0$
  ($S$-wave) and total angular momentum
  $J=1$. There is a small admixture of $D$-waves as well, leading to the
  following coupled differential equations for the deuteron
  \begin{eqnarray}
   &-\dfrac{\hbar^2}{M}\dfrac{\diff^2u(r)}{\diff r^2}+V_\mathrm{C} u(r)
    +\sqrt{8}V_\mathrm{T} w(r)=E_d u(r),
    \notag \\
   &-\dfrac{\hbar^2}{M}\dfrac{\diff^2w(r)}{\diff r^2}+
    \left(\dfrac{6\hbar^2}{Mr^2} +
     V_\mathrm{C} - 2V_\mathrm{T} -3V_\mathrm{LS}\right)w(r)
    \notag \\
   &+\sqrt{8}V_\mathrm{T} u(r)=E_d w(r),\label{eq:deuteron}
  \end{eqnarray}
  where $u(r)$ and $w(r)$ are the radial wavefunctions of the $S$-wave
  and the $D$-wave, respectively. The potentials
  $V_\mathrm{C},V_\mathrm{LS}$ and $V_\mathrm{T}$ are the central, spin-orbit
  and tensor forces, respectively. Knowing the solution of
  Eq.~(\ref{eq:deuteron}), we can integrate out the $D$-wave degrees of
  freedom and obtain the following effective central force 
  \begin{eqnarray}
   {\label{eq:Veff_central}}
    \Veff(r;\TSO) = V_\mathrm{C}(r;\TSO) + \Delta \Veff(r;\TSO), \notag \\
   \Delta \Veff(r;\TSO) \equiv \sqrt{8}V_\mathrm{T}(r)\dfrac{w(r)}{u(r)}.
  \end{eqnarray} 
  The effective central force $\Delta \Veff$ is comparable to
  $V_\mathrm{C}$ in strength and it makes the
  $\TSO$ channel
  the most attractive one~\cite{PhysRevC.74.034330,PTPS.39.23}.
  This effective central force makes the deuteron bound 
  for the $\TSO$ channel.
  We can regard this equation as a special case of
  Eq.~(\ref{eq:general}). The effective central force comes from a
  second-order effect due to tensor force, since both the initial
  and the final state have orbital angular momentum $0$. As a
  consequence, the effective interaction for the $T=0$ channel is
  enhanced by the  renormalization procedure due to the short-range part of
  the tensor force. This  is however not the case in $T=1$ channel. It
  reflects the property of the deuteron, which is the only bound
  two-nucleon system. A similar mechanism may also explain the strong 
  cutoff dependence of the $\Vlowk$ interaction seen in the $T=0$ channel.
  
 \section{Tensor force in effective interaction for the shell model}
 \label{sec:veffsm}
 We discuss here the tensor-force component in the
 effective interactions for the shell model, using the decomposition of
 Eq.~(\ref{eq:dec}). We have calculated effective interactions for the
 shell model ($\Veffsm$) using many-body perturbation theory (MBPT) by
 considering the $\hat{Q}$-box 
 up to second and third order with
 folded diagrams included as well, starting from a renormalized $\Vlowk$
 interaction. The cutoff value used in the $\Vlowk$ calculation is set
 to $\Lambda=2.1~\fmi$~\cite{PhysRevC.70.061002}. 
 The model space ($P$-space) is chosen  to be the full $sd$-shell or the
 full $pf$-shell. In the construction of the $\Vlowk$ interaction, we
 renormalize the strong short-range repulsion of the NN interaction,
 and in MBPT we include further effects of truncations of the model
 space. The $\hat{Q}$-box is calculated by considering valence-linked 
 and connected diagrams with unperturbed single particle energies of the
 harmonic oscillator. The oscillator energy $\hbar \omega$ is set to be
 $14~\mathrm{MeV}$ and $11~\mathrm{MeV}$ for the $sd$-shell  and the
 $pf$-shell effective interactions, respectively. Degenerate
 perturbation theory is employed in  constructing the  effective
 interactions. 

 Since the $Q$-space is defined as the complement of the $P$-space,
 intermediate states arising in each diagram should be taken up to
 infinitely high oscillator shells. In our case, using a low-momentum
 interaction $\Vlowk$ with $\Lambda=2.1~\fmi$, full convergence of the
 monopole part of $\Veff$ is obtained with approximately $8
 \mathrm{-}10~\hbar\omega$ excitations in each diagram which makes up
 the $\hat{Q}$-box.
 \begin{figure*}[htb]
  \begin{center}
   \begin{tabular}{cc}
    \resizebox{80mm}{!}{\includegraphics[angle=270]
    {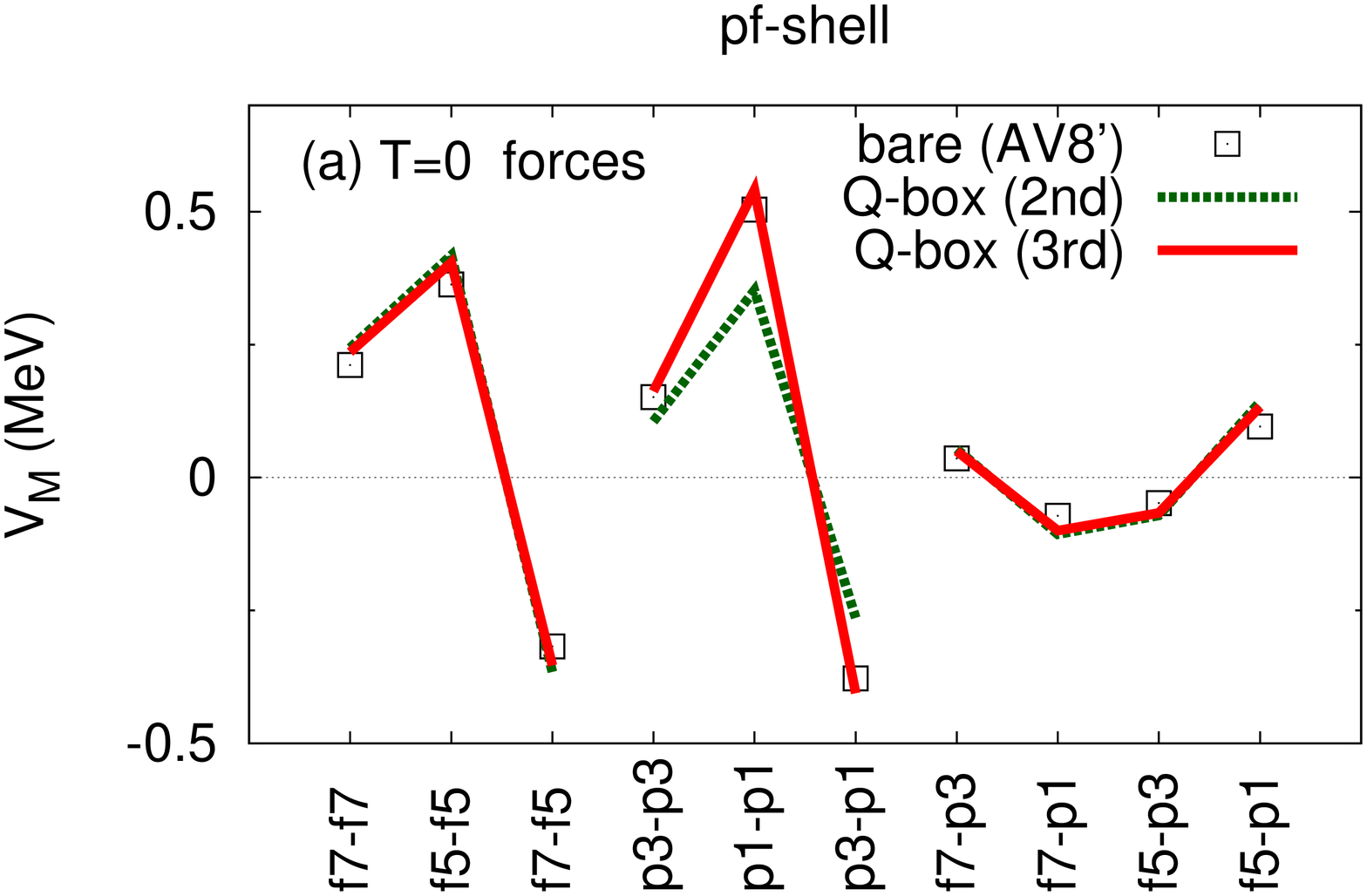}} &
    \resizebox{72.25mm}{!}{\includegraphics[angle=270]
    {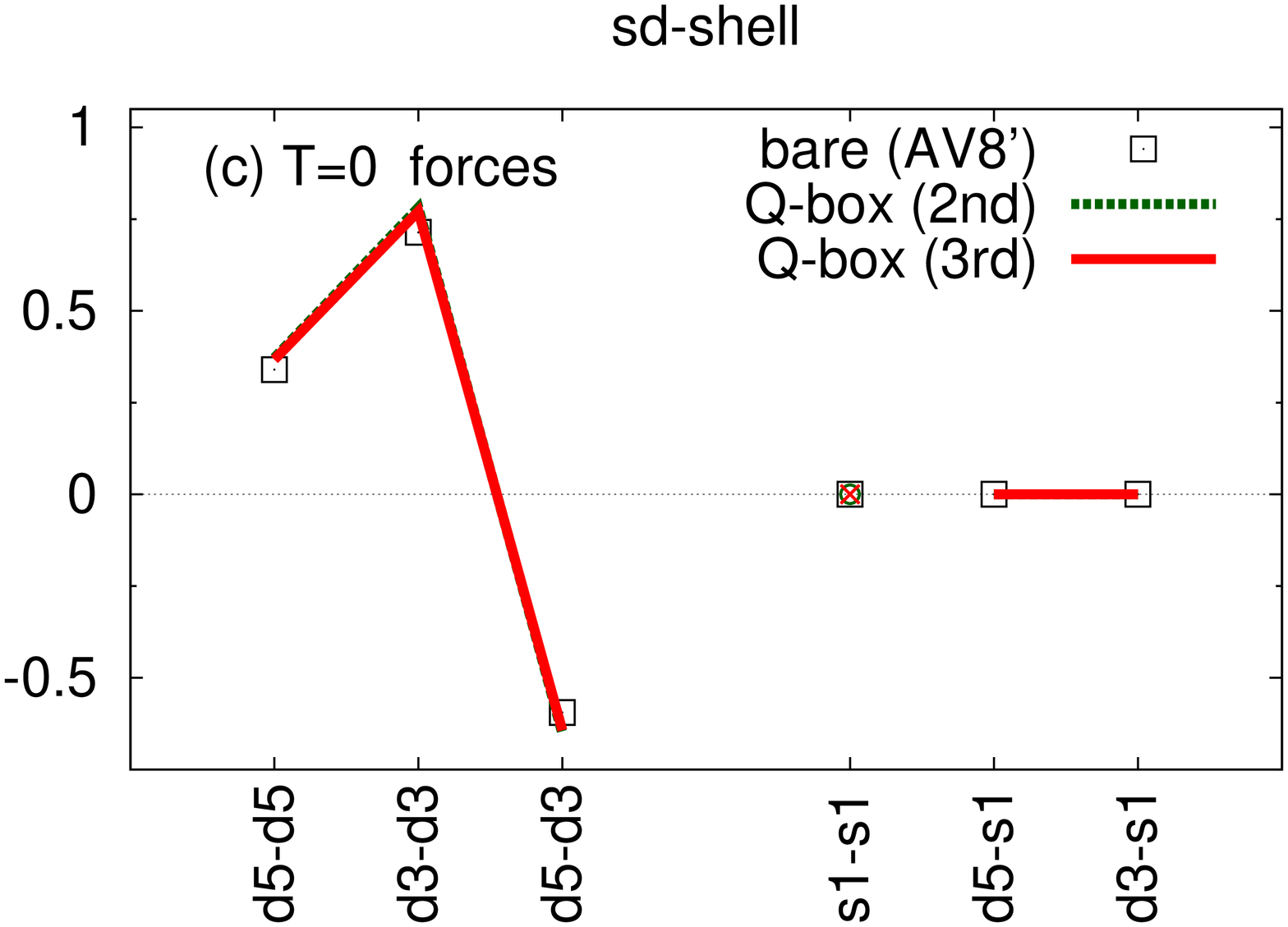}} \\
    \resizebox{80mm}{!}{\includegraphics[angle=270]
    {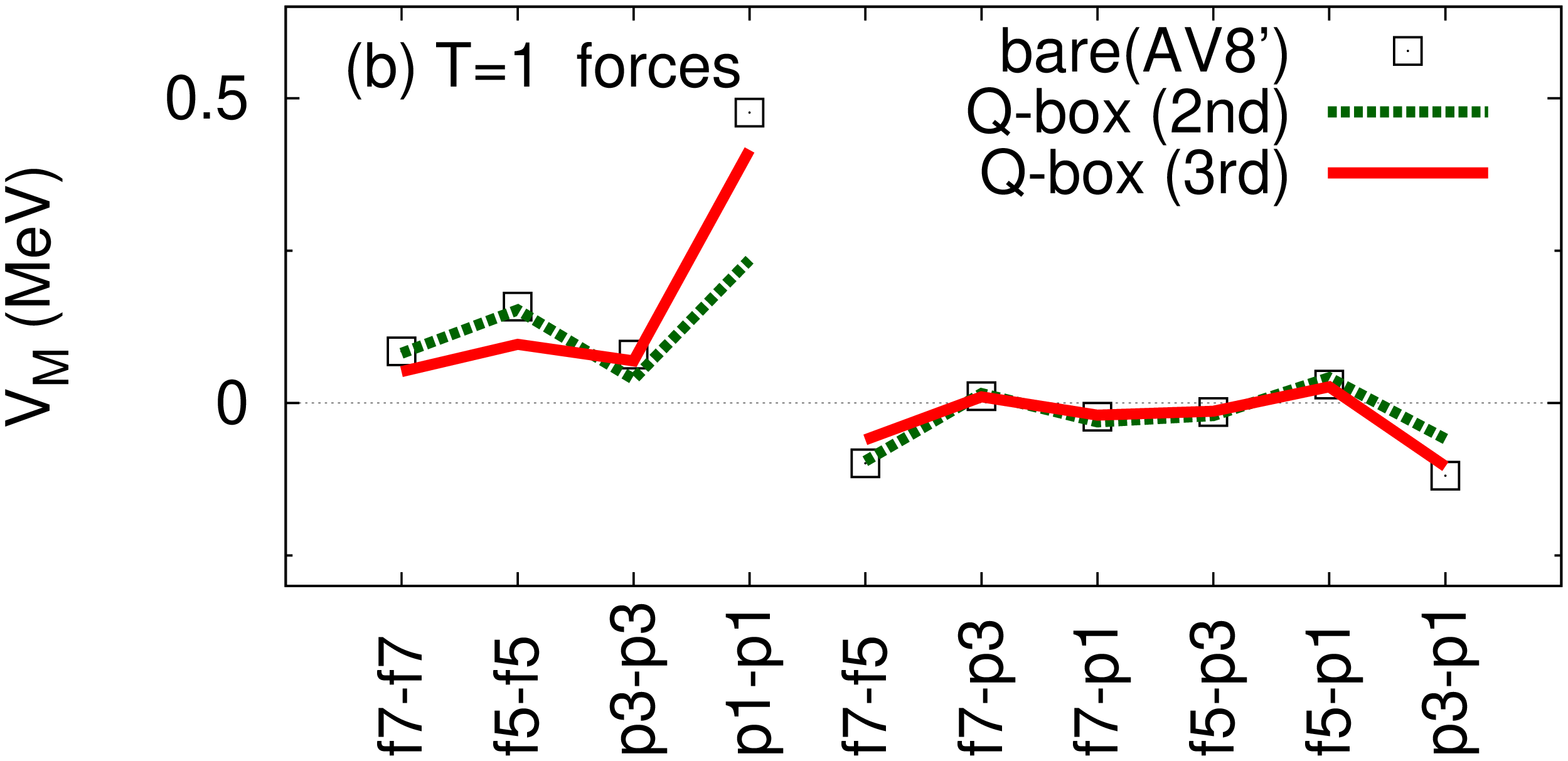}} & 
	\resizebox{72.25mm}{!}{\includegraphics[angle=270]
	{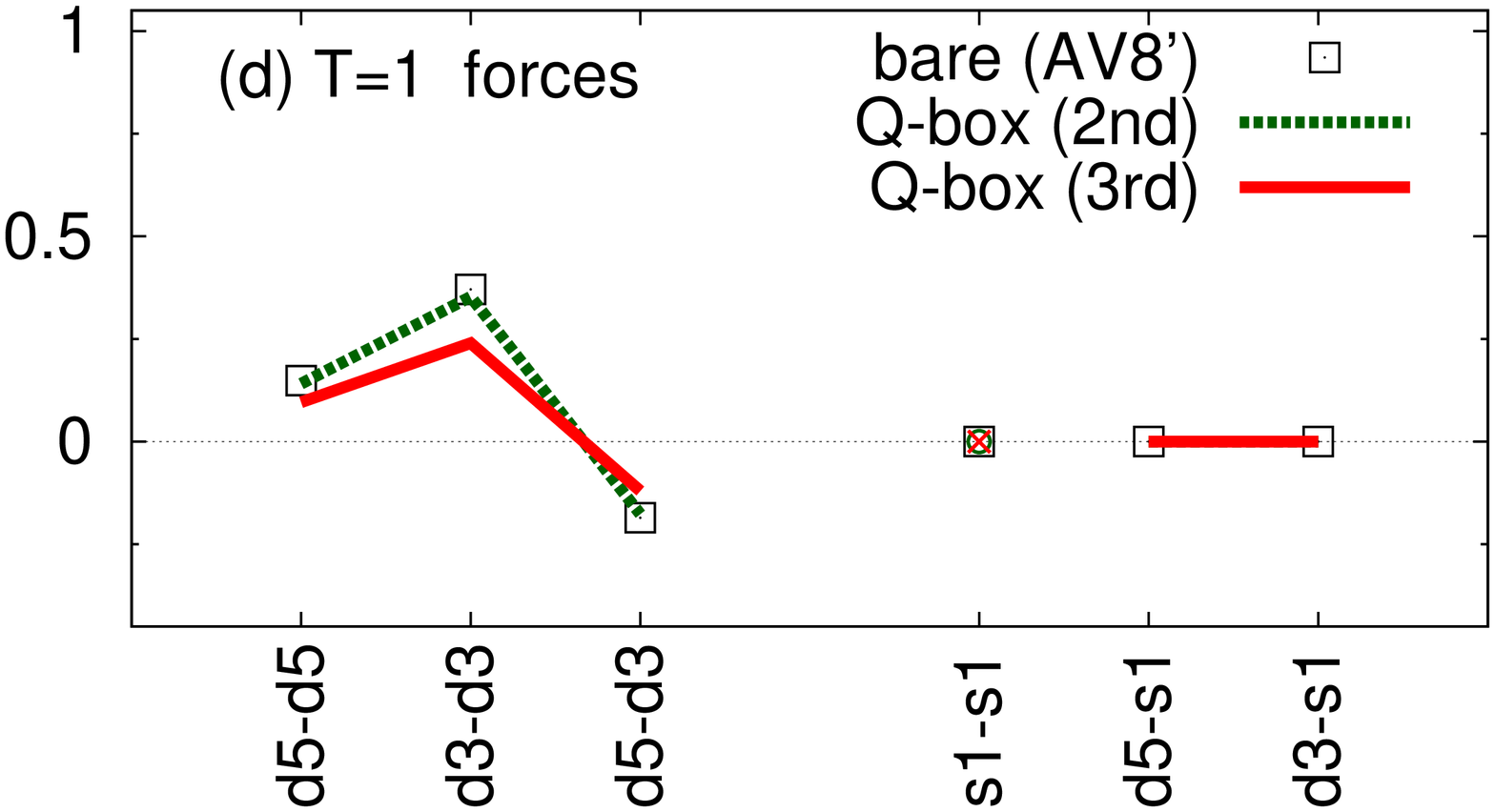}}\\
   \end{tabular}
   \caption{(color online.)
   The tensor force
   monopole component of the effective interaction for the shell model
   obtained by the $\hat{Q}$-box expansion to second and third order in
   the interaction, starting from $\Vlowk (\Lambda = 2.1~\fmi)$. The
   tensor-force component is obtained using the decomposition of
   Eq.~(\ref{eq:dec}).
   (a) $T=0$ forces in $pf$-shell, (b) $T=1$ forces in
   $pf$-shell, (c) $T=0$ forces in $sd$-shell and (d) $T=1$ forces in
   $sd$-shell.} 
    \label{fig:Qbox_ten}
  \end{center}
 \end{figure*}
 
 Figure~\ref{fig:Qbox_ten} shows the monopole part of the tensor force
 of $\Veffsm$ defined for the $sd$-shell or the $pf$-shell. As a general
 trend, one can see again that the monopole part of the tensor force of
 $\Veffsm$ fulfills our R-Persistency hypothesis to a good extent both in the
 $sd$-shell and in the $pf$-shell. Since the first order $\hat{Q}$-box
 is just the $\Vlowk$ interaction, 
 the results mean that the monopole part of the
 tensor force is dominated by the first-order term in the \Qbox~and the
 contributions from second or higher-order terms are 
 remarkably small. 
 These results can be understood by considering the specific angular
 momentum structure of the tensor force, which is a scalar product of
 two rank $2$ tensors in spin and coordinate spaces. In a perturbative
 correction to second or higher-order, such a complicated structure is
 smeared out and the resulting interaction consists mainly of a central
 force contribution. Therefore, as for the tensor-force component in the
 monopole interaction, it is the first-order contribution which is the
 dominant one.

To elucidate why higher-order terms in many-body perturbation theory are small, we consider
as an example a contribution from second order in the interaction, by far the largest higher-order term.

 The Hamiltonian causing the present second-order perturbation can be written as
 \begin{equation}
  \label{eq:H1}
   H_1 =
    \sum_{p=0,1,2}
   w_p (U^{(p)} \cdot  X^{(p)}),
 \end{equation}
 where $w_p$ represents an interaction strength, $U^{(p)}$ and $X^{(p)}$ 
 are operators of rank $p$ in spin space and coordinate
 space, respectively.  
A contribution from second-order in perturbation theory 
to a state
 $\phi$ can then be written as
 \begin{equation}
 \label{eq:2ndPT}
   \eta (\phi) = - \sum_{j} \frac{\bra{\phi} H_1 \ket{\psi_j}
                   \bra{\psi_j} H_1 \ket{\phi} }{\Delta E_j},
 \end{equation} 
 where $\psi_j$ defines an  intermediate state with 
 energy denominator $\Delta E_j$.    
 The summation is done over all  intermediate states $\psi_j$.
 As far as $\psi_j$ varies in this summation, within a fixed 
 configuration with respect to harmonic-oscillator (HO) 
 shells, $\Delta E_j$ remains constant due to the degeneracy of  
 single-particle energies in a given HO shell. We mention that
 the usage of non-degenerate perturbation theory yields only small changes.
  
 Such a configuration for a given  HO shell is denoted by $S$.
 As $\Delta E_j$ is a constant within a fixed shell $S$, we label it 
 as $\Delta E_S$.
 Note that $S$ corresponds to a part of the $Q$-space, while 
 $\phi$ is in the $P$-space.  
The term $\eta (\phi)$ can then be decomposed into contributions from 
 individual $S$'s as
 \begin{equation}
 \label{eq:2ndPT1}
   \eta (\phi) = - \sum_{S} \frac{\zeta (\phi, S)}{\Delta E_S},
 \end{equation} 
 where
 \begin{equation}
 \label{eq:2ndPT2}
   \zeta (\phi, S) = \sum_{j \in S} \bra{\phi} H_1 \ket{\psi_j}
                   \bra{\psi_j} H_1 \ket{\phi}.
 \end{equation} 
 For a given $S$, all $\psi_j$s are included, and the summation can be
 replaced by the closure relation as
 \begin{equation}
 \label{eq:2ndPT3}
   \zeta (\phi, S) = \bra{\phi} \{ H_1 H_1 \}_S \ket{\phi},
 \end{equation} 
 where the parentheses $\{ \, \}_S$ are introduced to indicate that
 the second $H_1$ changes $\phi$ to an $S$-configuration state 
 in the $Q$-space and the first $H_1$ moves it back to state $\phi$ in the
 $P$-space.  In other words, $H_1 H_1$ in this equation cannot be 
 a simple product, but a certain contraction is needed as we
 shall show soon.

 By utilizing Eq.~(\ref{eq:H1}), we obtain
 \begin{widetext}
  \begin{eqnarray}
  \label{eq:UUXX}
   \{ H_1 H_1 \}_S
   &= &\sum_{p_1,p_2} w_{p_1}w_{p_2} 
       \{(U^{(p_1)} \cdot X^{(p_1)})(U^{(p_2)} \cdot X^{(p_2)})\}_S 
      \notag \\ 
   &= &\sum_{k=0,1,2}(2k+1) \left( \sum_{p_1,p_2} w_{p_1}w_{p_2}  
    \9j {p_1}{p_2}k{p_1}{p_2}k000 
    \{\left[[U^{(p_1)} \times U^{(p_2)}]^{(k)} \times
     [X^{(p_1)} \times X^{(p_2)}]^{(k)} \right]^{(0)} \}_S \right) ,
  \end{eqnarray}
 \end{widetext}
 where the terms in curly brackets are $9j$ symbols and $k$ implies the rank of the 
recoupling.
 The operator $\{ [U^{(p_1)} \times U^{(p_2)}]^{(k)} \}_S$ acts 
 in the $P$-space as a rank-$k$ two-body operator in spin space, 
 while $\{ [X^{(p_1)} \times X^{(p_2)}]^{(k)} \}_S$ 
 acts as a rank-$k$ two-body operator in coordinate space.
 Because the contraction due to the elimination of the $Q$-space
 does not affect the angular momentum properties, 
 the variable $k=0,1,2$ represents induced central, 
 spin-orbit and tensor forces in the $P$-space, respectively.

 Since we are mainly interested in
 the tensor component, we focus on the case of $k=2$, with the obvious
 restriction $p_1+p_2 \geq 2$.  
 Since the above $9j$ symbols is proportional to $1/\sqrt{(2p_1+1)(2p_2+1)}$, 
 it is easy to convince oneself that the central force component
 receives the largest contribution from the $9j$ symbol. Furthermore,
 for our analyses 
 it is important to keep in
 mind that  the expectation value of the 
central component is the largest in absolute value, the tensor component
 the second largest and the spin-orbit term gives rise to the
 smallest contribution to the renormalized  $\Vlowk$ interaction. 

 From these considerations, for $k=2$ the largest contribution comes
 from the combination $p_1=0,~p_2=2$ or $p_1=2,~p_2=0$ in
 Eq.~(\ref{eq:H1}), that is either a  central-tensor or a tensor-central
 combination.
 Let us now discuss this case.  We assume without loss of generality 
 that the tensor component of $H_1$ acts on the ket state of the
 matrix element being considered.  While the central component of $H_1$ 
 acts afterward on this state, we can also consider that this central 
 force acts to the left on the bra state.  We then take the overlap 
 between these two states by considering
 one by the tensor on the ket side, and the other by the central on the bra side.
 These two states are sum-rule states for the two forces
 within the $S$-configuration space.  As the
 central force and the tensor force are very different in nature,
 such sum-rule states are very different from each other in general,
 leading to a very small overlap.  This is the main reason
 why the combination of the central force and the tensor force produces small  
contributions.  
 
 This argument does not hold for the case where the tensor component 
 of $H_1$ acts
 twice in the term to second-order in perturbation theory.  
However, due to the angular
 momentum coupling, the product of two tensor forces 
 ($p_1=p_2=2$ in Eq.~(\ref{eq:H1})) yield small 
 contributions to the $k=2$ terms of Eq.~(\ref{eq:H1}).
 For higher orders, other tensor-force components may show up, but there is
 no mechanism to enhance their contributions.
 
 The small contribution of the tensor force in MBPT can be viewed 
 to be reasonable also under the following intuitive 
 picture: after multiple actions of the forces, the spin
 dependence is smeared out, and only the
 distance between two interacting nucleons becomes the primary factor to
 the whole processes.
 This results in the dominance of the induced effective interaction 
 by the central components and yields only a  minor
 change in the tensor component.

It is instructive to study in more detail the contributions to
second-order in perturbation theory.
To do so, we single out the by far largest
second-order term, namely the so-called core-polarization term,
depicted as diagram (c) in Fig.~\ref{fig:qbox_2nd}.
For the core-polarization diagram we can show
 that the contribution to the tensor force vanishes by simple angular
 momentum algebra arguments.
 The contribution to a specific core-polarization matrix element can then be written as 
 \begin{widetext}
  \begin{eqnarray}
   \bra{a m_a b m_b} V^{\mathrm{cp-eff}}_T \ket{c m_c d m_d}
    &= &\sum_{p,m_p,h,m_h}
    \bra{a m_a p m_p}V_C\ket{c m_c h m_h}
    \bra{h m_h b m_b}V_T\ket{p m_p d m_d}
    /\Delta E \notag\\
   &= &\sum_{n_p,l_p,n_h,l_h} 
    \Bigl( \sum_{j_p,m_p,j_h,m_h}\bra{a m_a p m_p}V_C\ket{c m_c h m_h}
    \bra{h m_h b m_b}V_T\ket{p m_p d m_d}\Bigr) 
    /\Delta E_S \notag\\
   &= &\sum_{n_p,l_p,n_h,l_h} 
    \Bigl( \sum_{m_{lp},m_{sp},m_{lh},m_{sh}}
    \bra{a m_a n_p l_p m_{lp} m_{sp}}V_C\ket{c m_c n_h l_h m_{lh} m_{sh}}
    \notag\\
   & &\times \bra{n_h l_h m_{lh} m_{sh} b
    m_b}V_T\ket{n_p l_p m_{lp} m_{sp} d m_d}\Bigr)
    /\Delta E_S,
    \label{eq:t_and_c}
  \end{eqnarray}
 \end{widetext}
 where $V^{\mathrm{cp-eff}}_T$ is the induced tensor force, 
 $V_T$ and $V_C$ are the tensor force and central force components from $H_1$, respectively. 
With a harmonic oscillator basis, the term $\Delta E_S$ represents an 
 energy denominator which is constant  for a given set of quantum numbers $n_p, n_h, l_p$ and $l_h$, 
 as discussed above. 
 Here $a=(n_a,l_a,j_a)$, and $m_a$
 denotes magnetic substate of $l_a$. Note that the two-body states are not
 antisymmetrized. The states $p$ and $h$
 represent particle and hole states, respectively. In the third
 line of the equation, only particle and hole states are transformed to 
 the $ls$ coupling scheme. Note that the intermediate states are summed
 up to fulfill spin-saturation within each HO major shell.
  
 We can divide the contribution into two different types according to
 the spin dependence of the central force. One comes from the terms
 whose central force part $V_C$ includes $\sigma\cdot\sigma$ (type I)
 and the other does not (type II). With our summation tailored to a
 spin-saturated core or an excluded $Q$-space with all spin-orbit
 partners, we can prove that type II contributions always vanish because
 the first factor is diagonal with respect to spin, that is
 $m_{sp}=m_{sh}$, and the second factor is zero when we sum over
 spin-saturated contributions. Therefore, only a spin-dependent central
 force results in non-vanishing contributions to the tensor force for
 higher-order terms in $\Veffsm$. Finally, the contribution to the
 tensor force from the spin-dependent central force is quite small
 because the spin-dependent central force is generally by far smaller
 than the spin-independent central force in modern realistic NN
 potentials.

 In conclusion, medium effects produce minor contributions to the
 tensor-force component, resulting in a tensor-force component that is
 dominated by the bare NN interaction. Our hypothesis about
 Renormalization Persistency
 is fulfilled to a good extent by the tensor
 force.

 Finally, although our analyses has been performed
 within one major shell only, one should note that this persistency of
 the tensor-force component via a MBPT renormalization should also hold
 in for model spaces which span several shells.

 \section{Two-body matrix elements of the tensor force}
 \label{sec:multipole}
 In this section we study further renormalization properties of the tensor force by
 including higher multipole components.
 \begin{figure*}[htb]
  \begin{center}
   \begin{tabular}{cc}
    \resizebox{80mm}{!}{\includegraphics[angle=270]
    {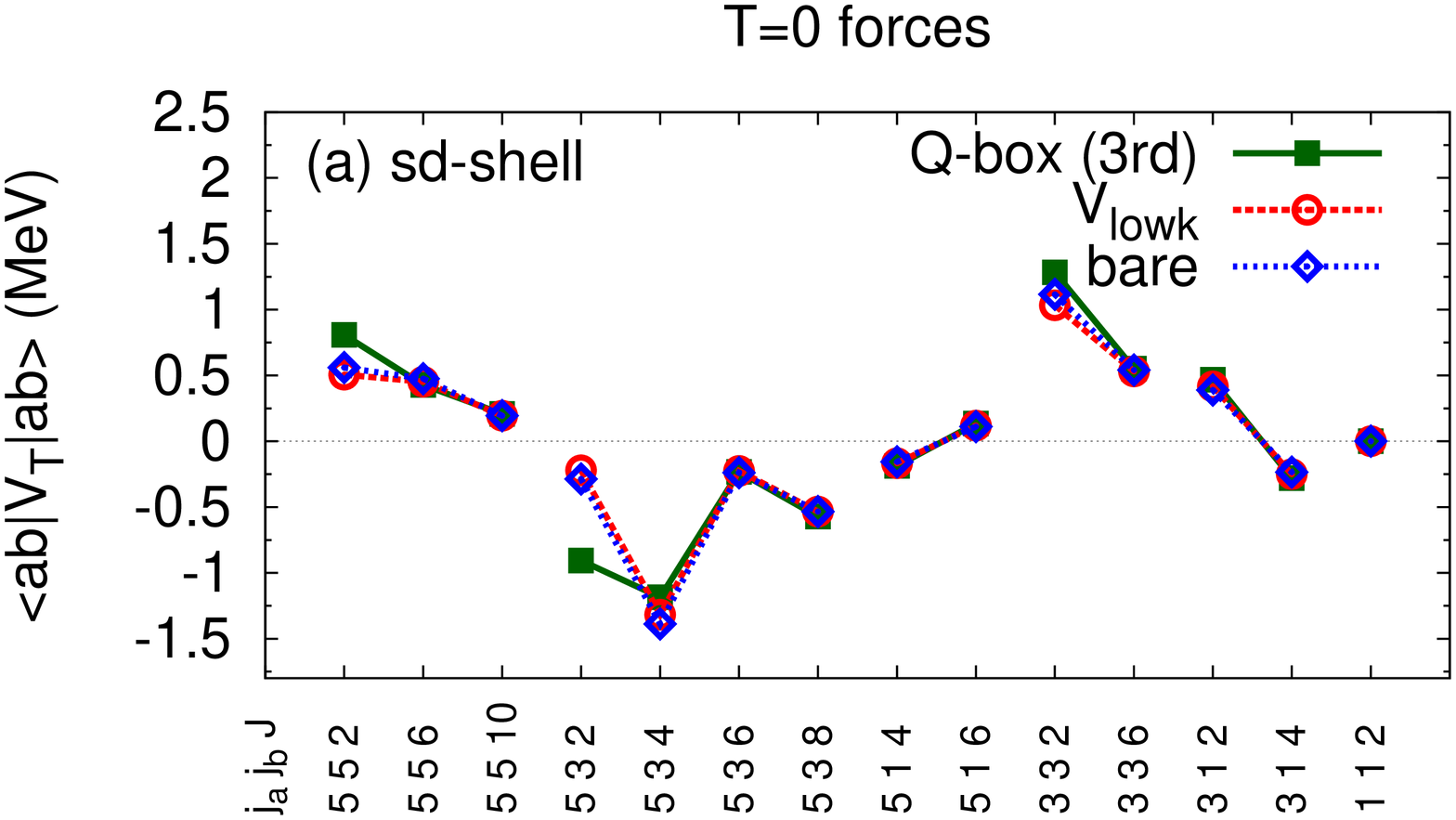}} &
    \resizebox{72.25mm}{!}{\includegraphics[angle=270]
    {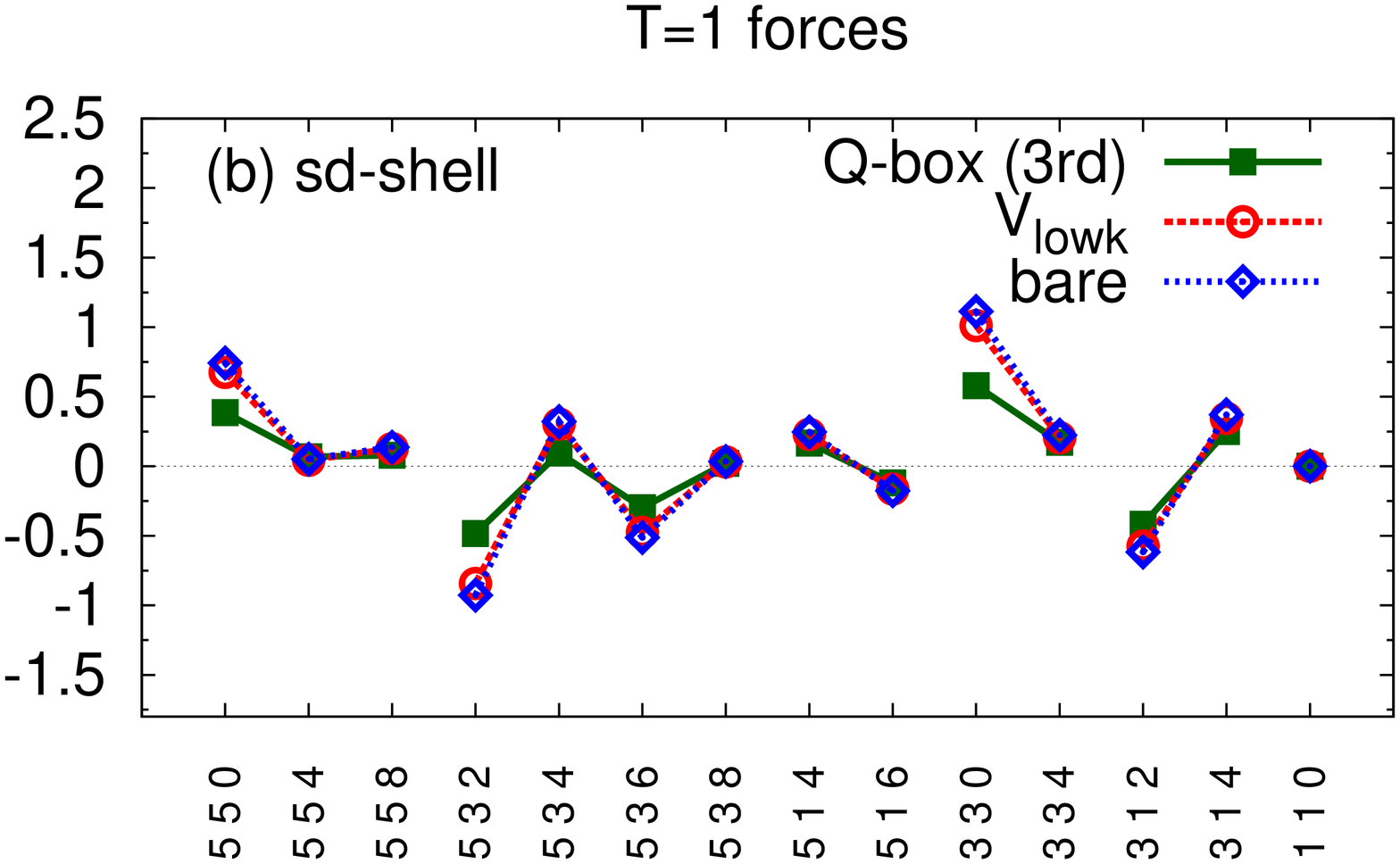}} \\
    \resizebox{80mm}{!}{\includegraphics[angle=270]
    {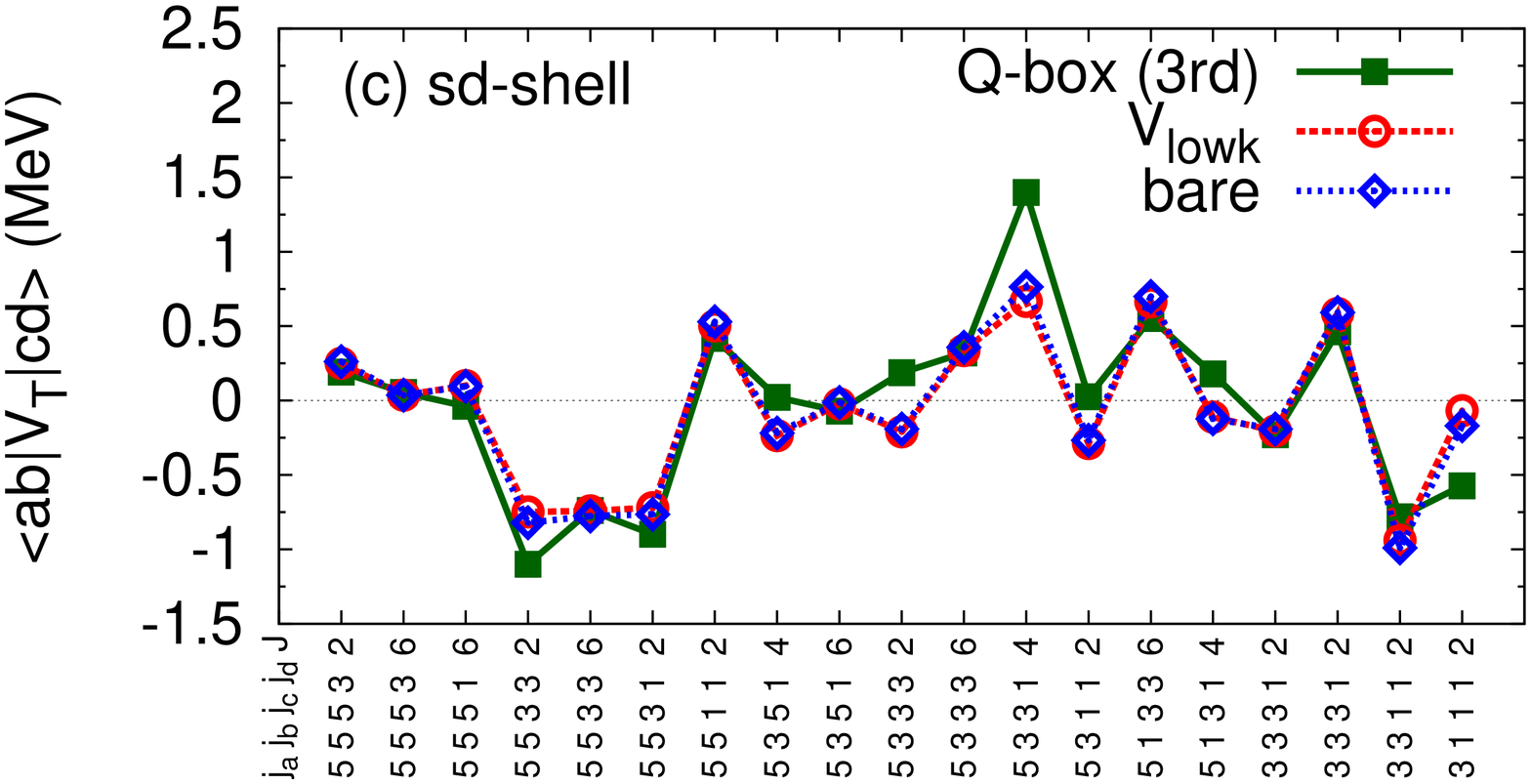}} &
    \resizebox{72.25mm}{!}{\includegraphics[angle=270]
	{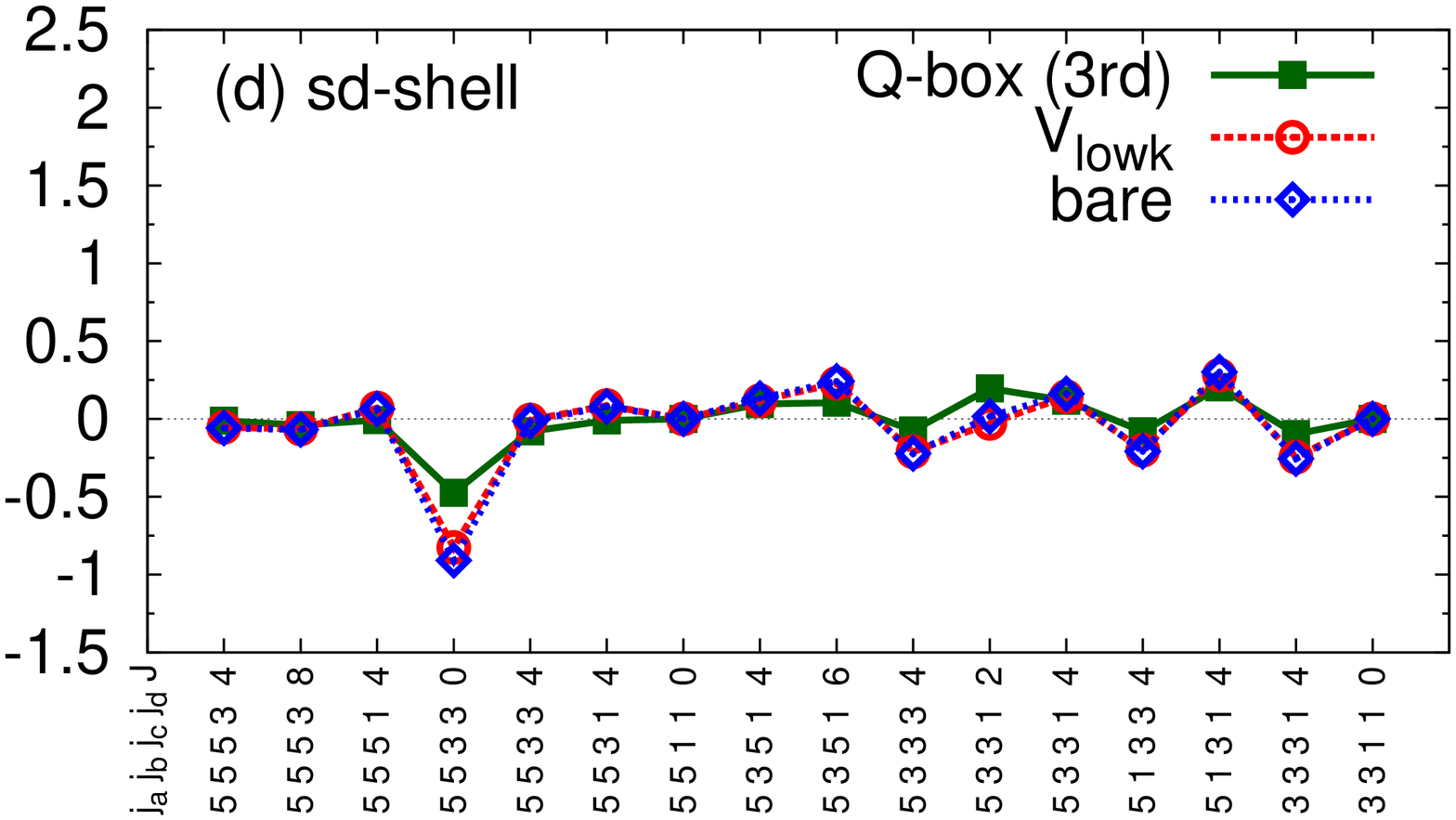}} \\
    \resizebox{80mm}{!}{\includegraphics[angle=270]
    {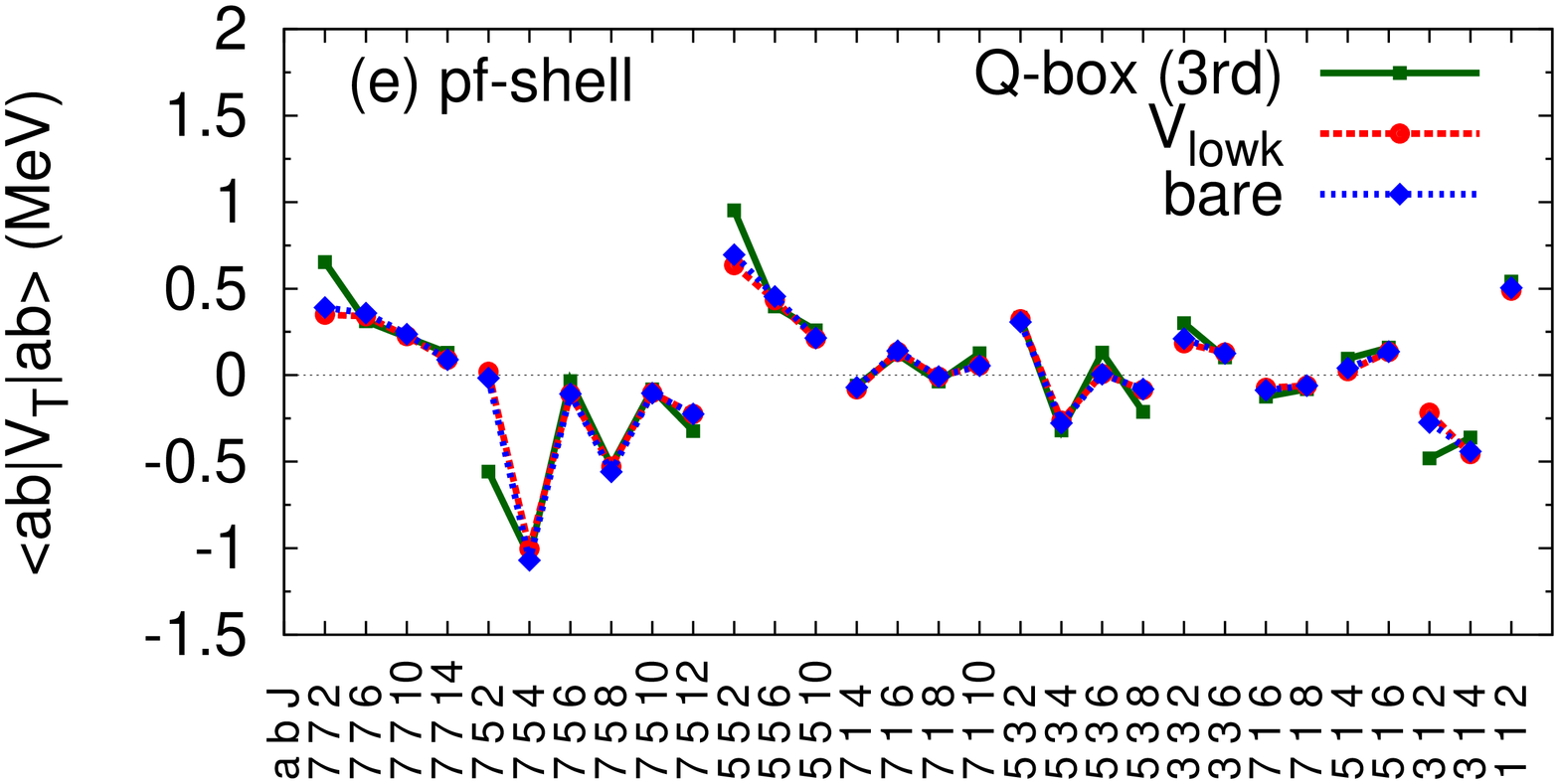}} &
    \resizebox{72.25mm}{!}{\includegraphics[angle=270]
	{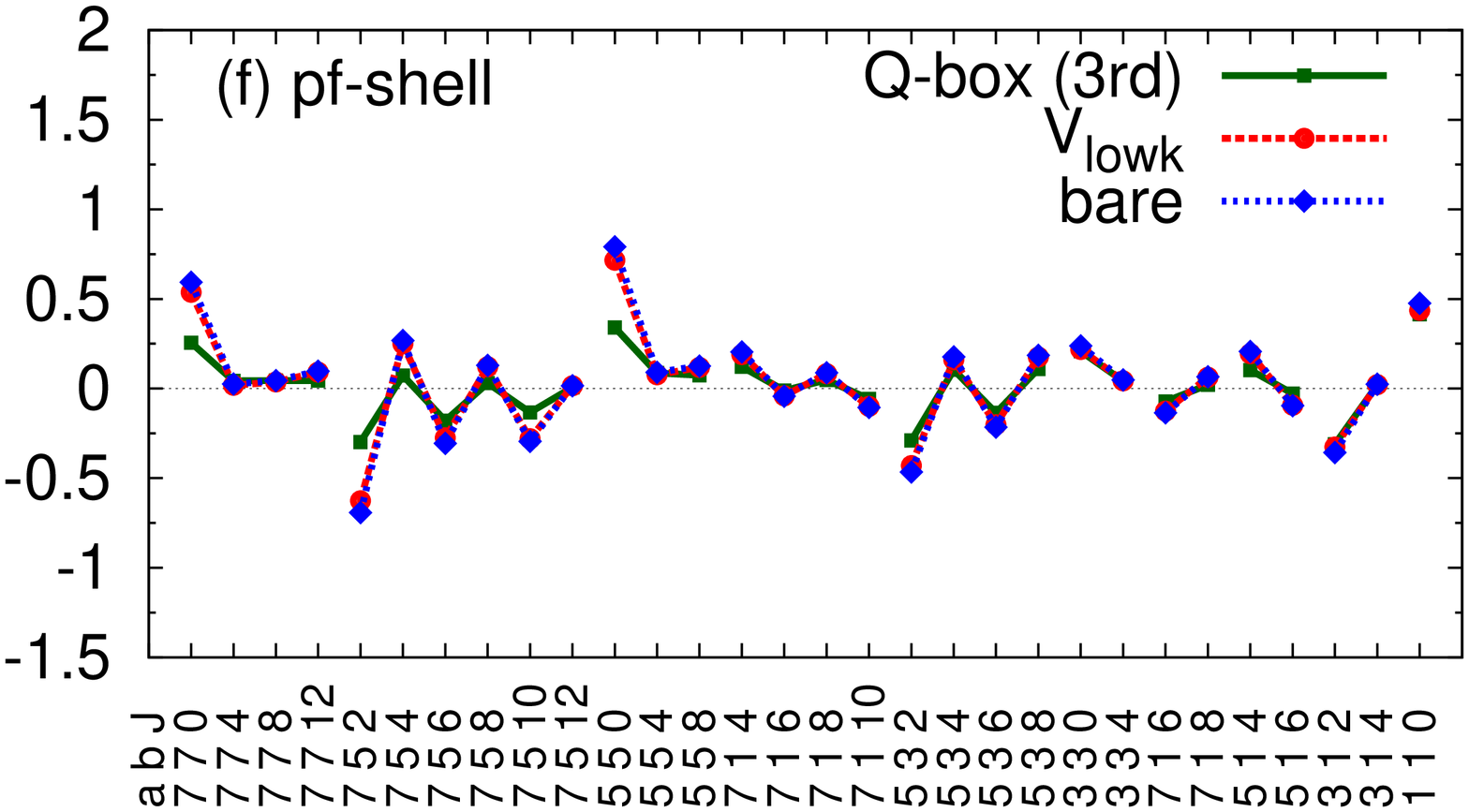}} \\
    \resizebox{80mm}{!}{\includegraphics[angle=270]
    {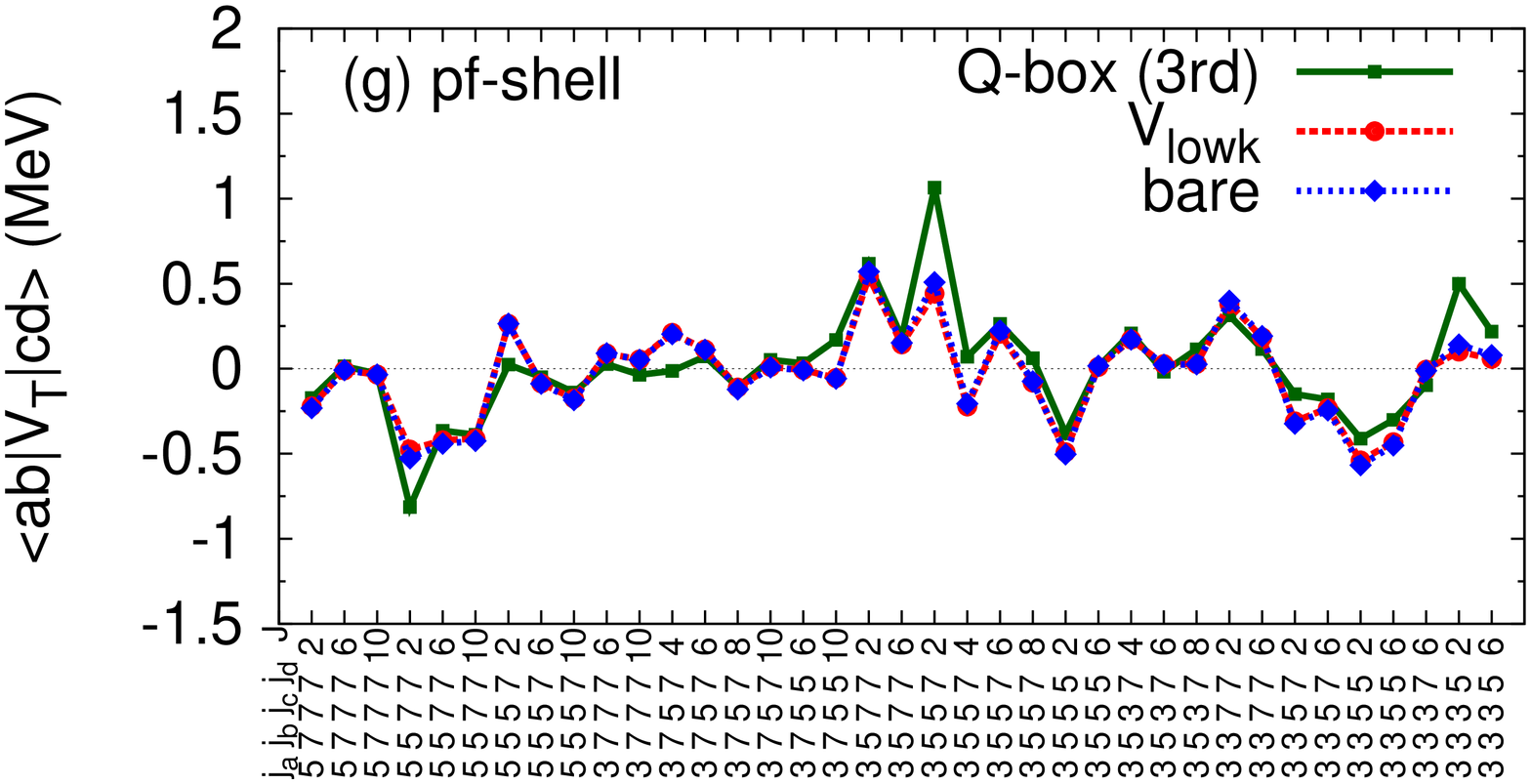}} &
	\resizebox{72.25mm}{!}{\includegraphics[angle=270]
	{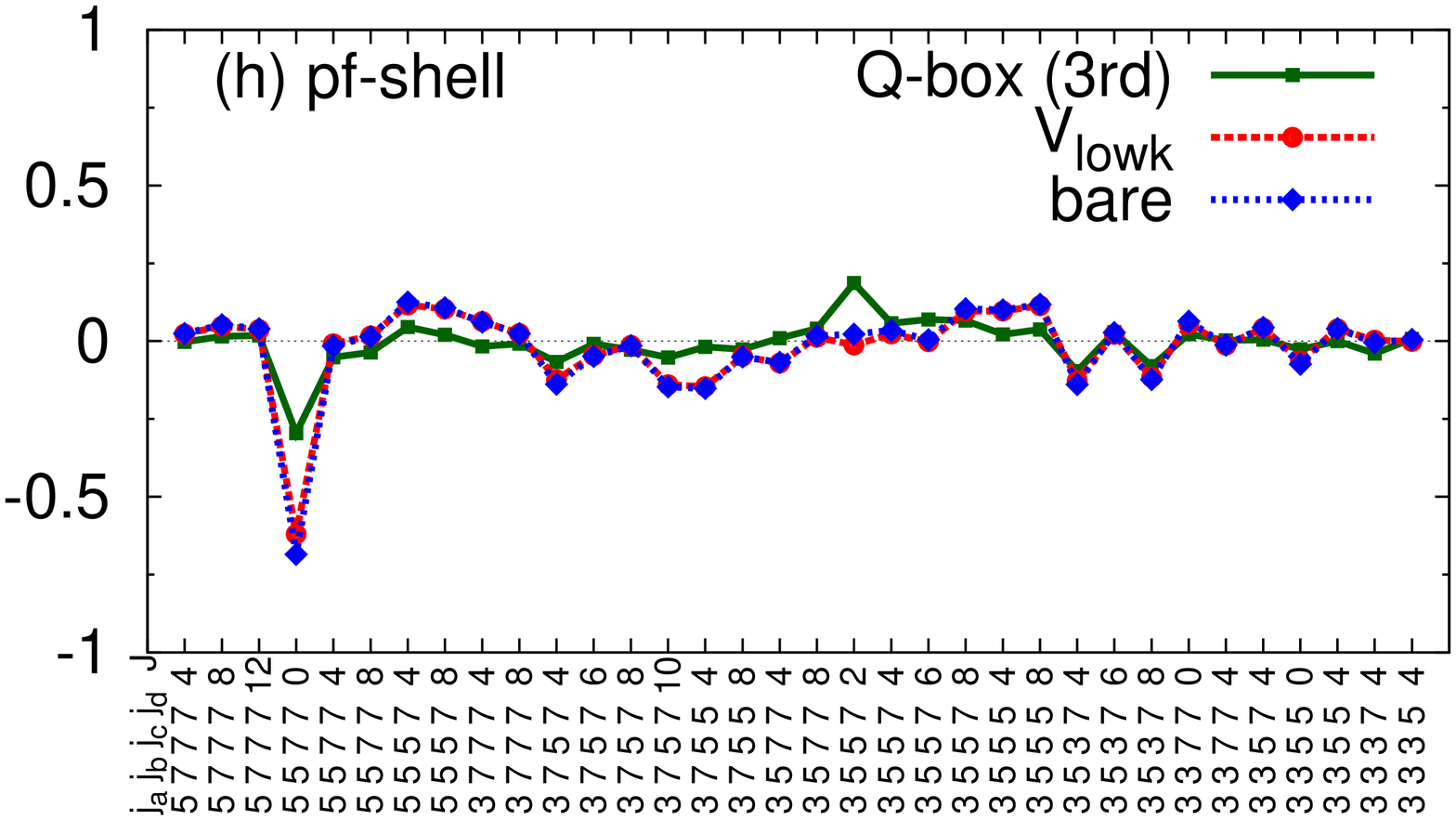}} \\
    \resizebox{80mm}{!}{\includegraphics[angle=270]
    {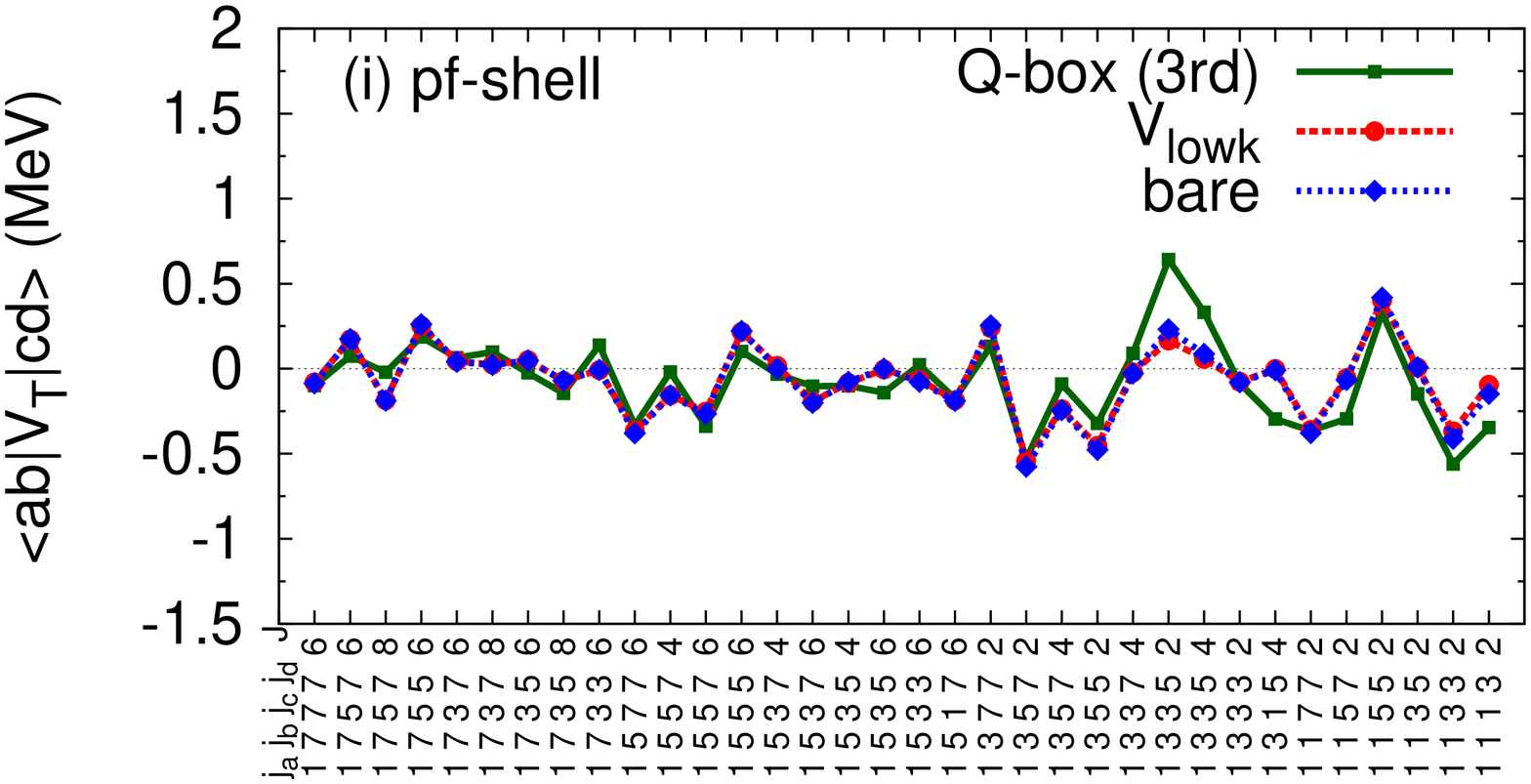}} &
	\resizebox{72.25mm}{!}{\includegraphics[angle=270]
	{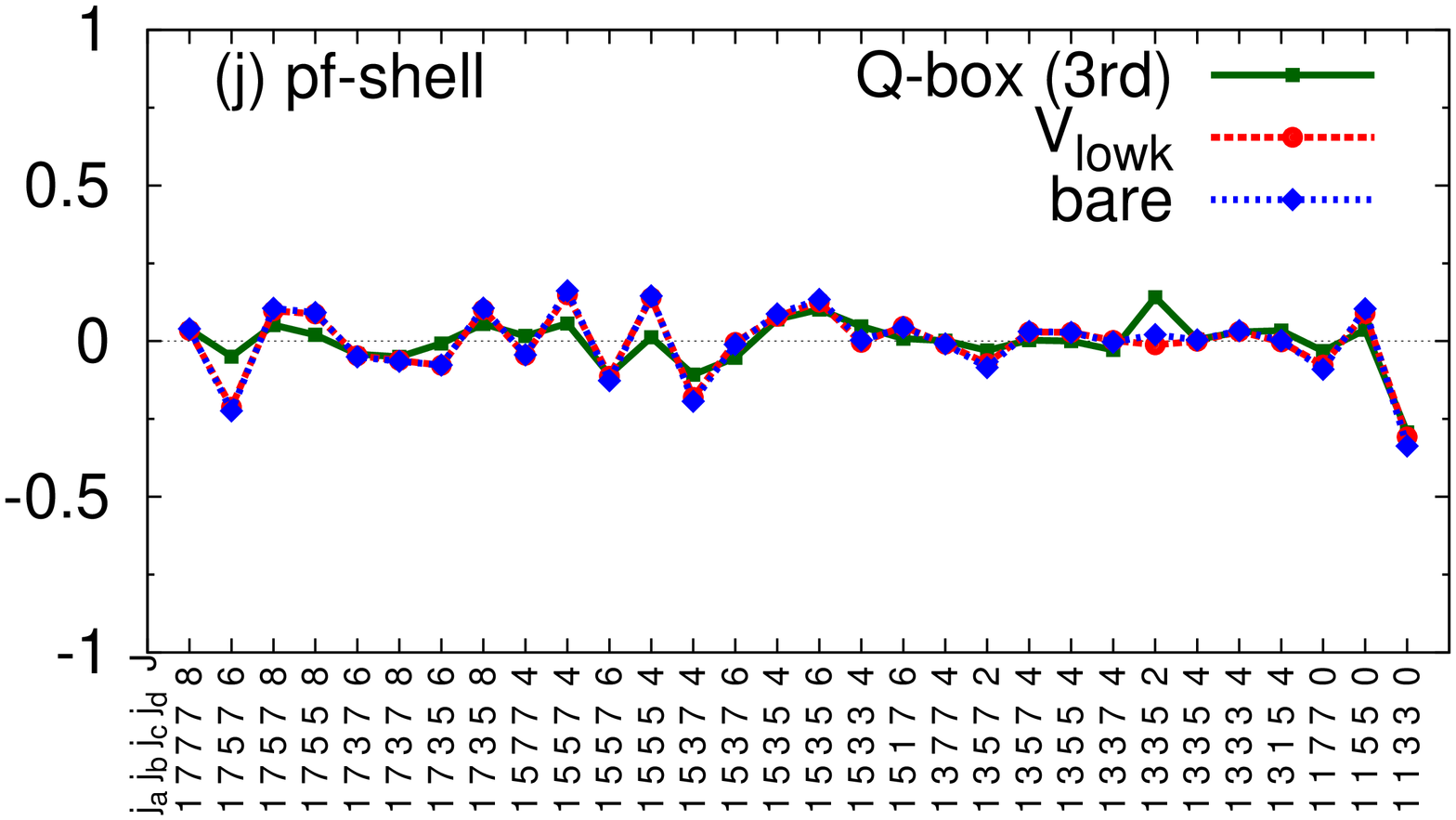}} \\
   \end{tabular}
   \caption{(color online.)
   Diagonal and non-diagonal matrix elements of the tensor force
   component from effective interactions using the  AV8' potential.}
   \label{fig:multi_av8_d}
  \end{center}
 \end{figure*}

 Figure~\ref{fig:multi_av8_d} shows the diagonal and non-diagonal
 matrix elements of the bare tensor force from the AV8' potential, the
 renormalized $\Vlowk$ interaction ($\Lambda=2.1~\fmi$) and $\Veffsm$
 obtained by the \Qbox~expansion up to the third order with folded diagrams to infinite order
 starting from AV8' interaction. This is similar to what was done
 in Figs.~\ref{fig:vlowk_ten} and \ref{fig:Qbox_ten}. In this figure, panels (a) to (d)
 stand for the $sd$-shell matrix elements. The diagonal matrix elements are shown in panels 
(a) and (b)  while the non-diagonal elements are shown in panels
 (c) and (d). Note that the diagonal matrix
 elements $\bra{j_a j_b}V\ket{j_a j_b}_{JT}$ are specified by the
 quantum numbers $j_a,~j_b$, and twice the total angular momentum $J$
 and total isospin $T$. The non-diagonal matrix elements $\bra{j_a
 j_b}V\ket{j_c j_d}_{JT}$ are specified by $j_a,~j_b,~j_c,~j_d,~J$ and
 $T$. The corresponding numbers for the $pf$-shell are shown in panels (e)
 through (j). In both the $sd$-shell and the $pf$-shell, the patterns
 are the same for all approaches to the effective interactions and thus the R-Persistency
 is approximately fulfilled. In particular, for the $\Vlowk$ renormalization procedure, 
we can hardly see
 any difference between the bare tensor force and the tensor force in
 the effective interaction $\Vlowk$. For the diagonal matrix elements,
 we can see small differences between the final \Qbox~and the bare
 tensor force, however, it does not contradict  the results with
 respect to the monopole component discussed above, mainly  because only
 matrix elements  with small values of the total  angular momentum display sizable
 differences.  Since the monopole terms are weighted by $2J+1$, matrix
 elements with larger values of the total angular momentum $J$ carry a
 much larger weight in Eq.~(\ref{eq:monopole}). In non-diagonal matrix 
 elements, we see somewhat larger differences.
 Their role in shell-model calculations needs to be investigated
 further. A spin-tensor analysis along these lines was made recently by
 Smirnova {\em et al}~\cite{Smirnova2010109}.

 \section{Analysis of other interaction models}
 \label{sec:n3lo}
 In the previous sections, we calculated effective interactions starting
 from the AV8' interaction, using a renormalized interaction  and many-body perturbation theory. 
 We found that the R-Persistency of the tensor force holds for all these
 renormalization procedures. An obvious question is whether or not the
 R-Persistency holds for other interaction models as well. In this
 section we address this issue as well.
 
\begin{figure*}[htb]
  \begin{center}
   \begin{tabular}{cc}
    \resizebox{80mm}{!}{\includegraphics[angle=270]
    {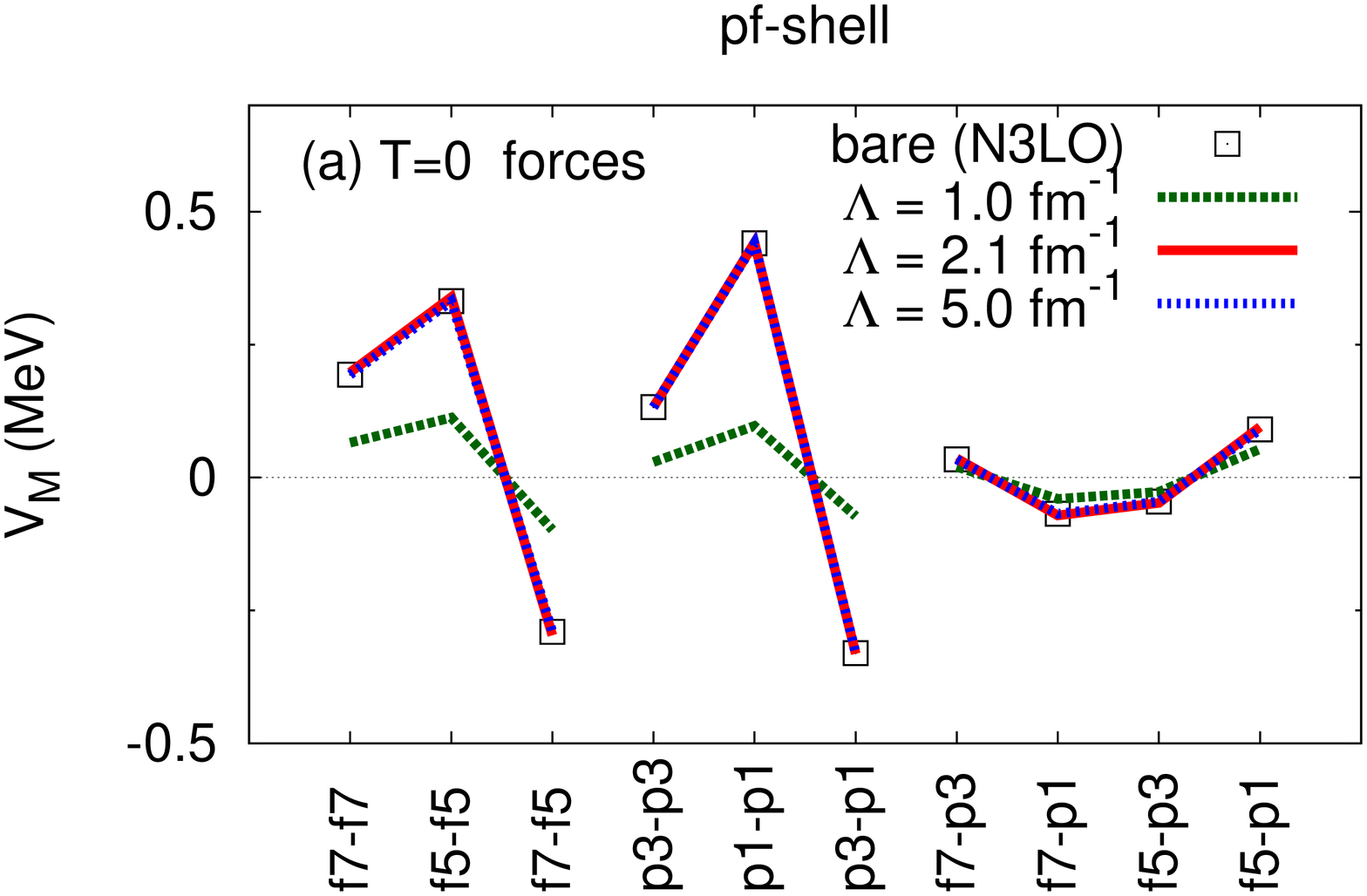}} & 
    \resizebox{72.25mm}{!}{\includegraphics[angle=270]
    {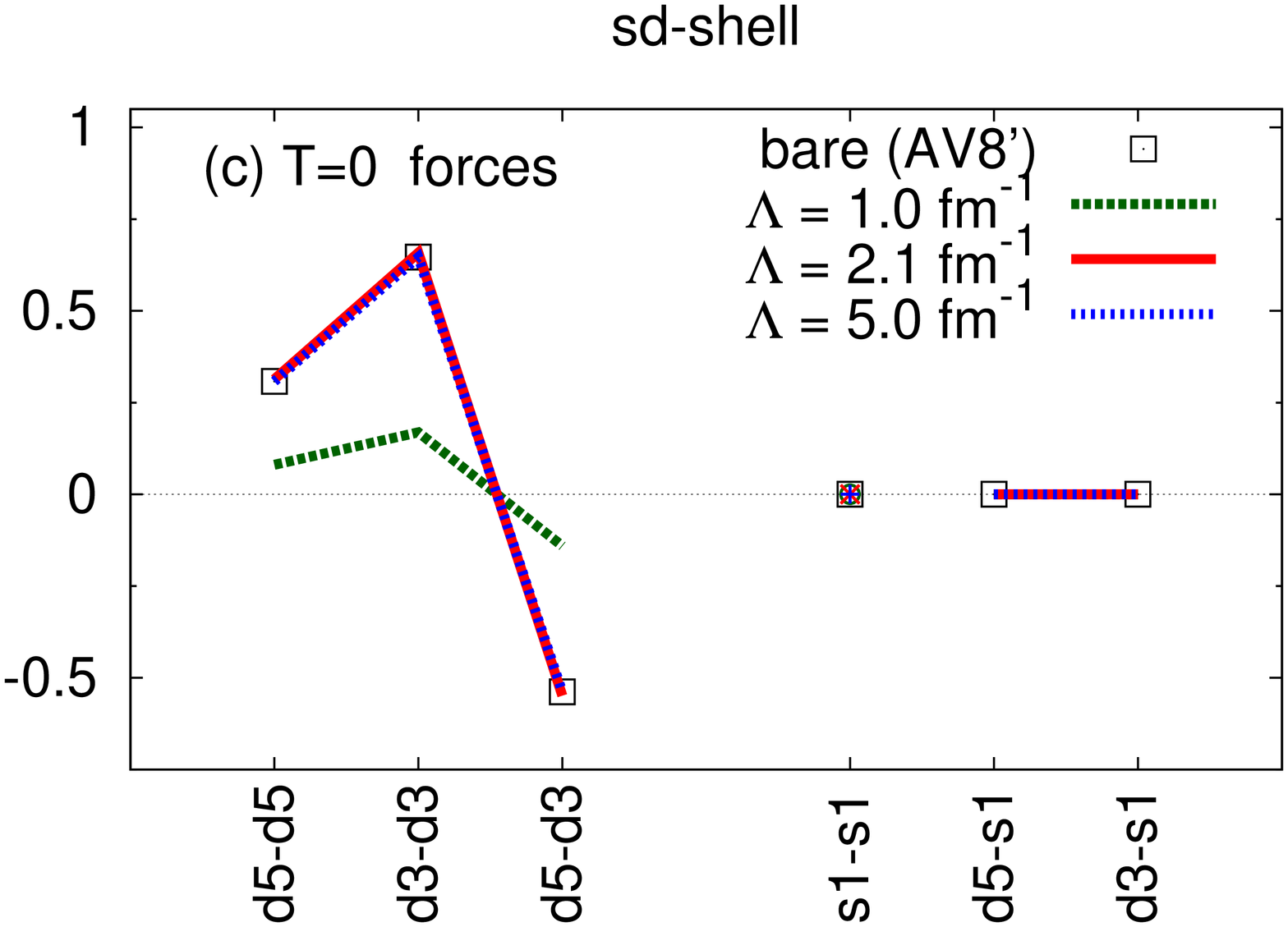}} \\ 
    \resizebox{80mm}{!}{\includegraphics[angle=270]
    {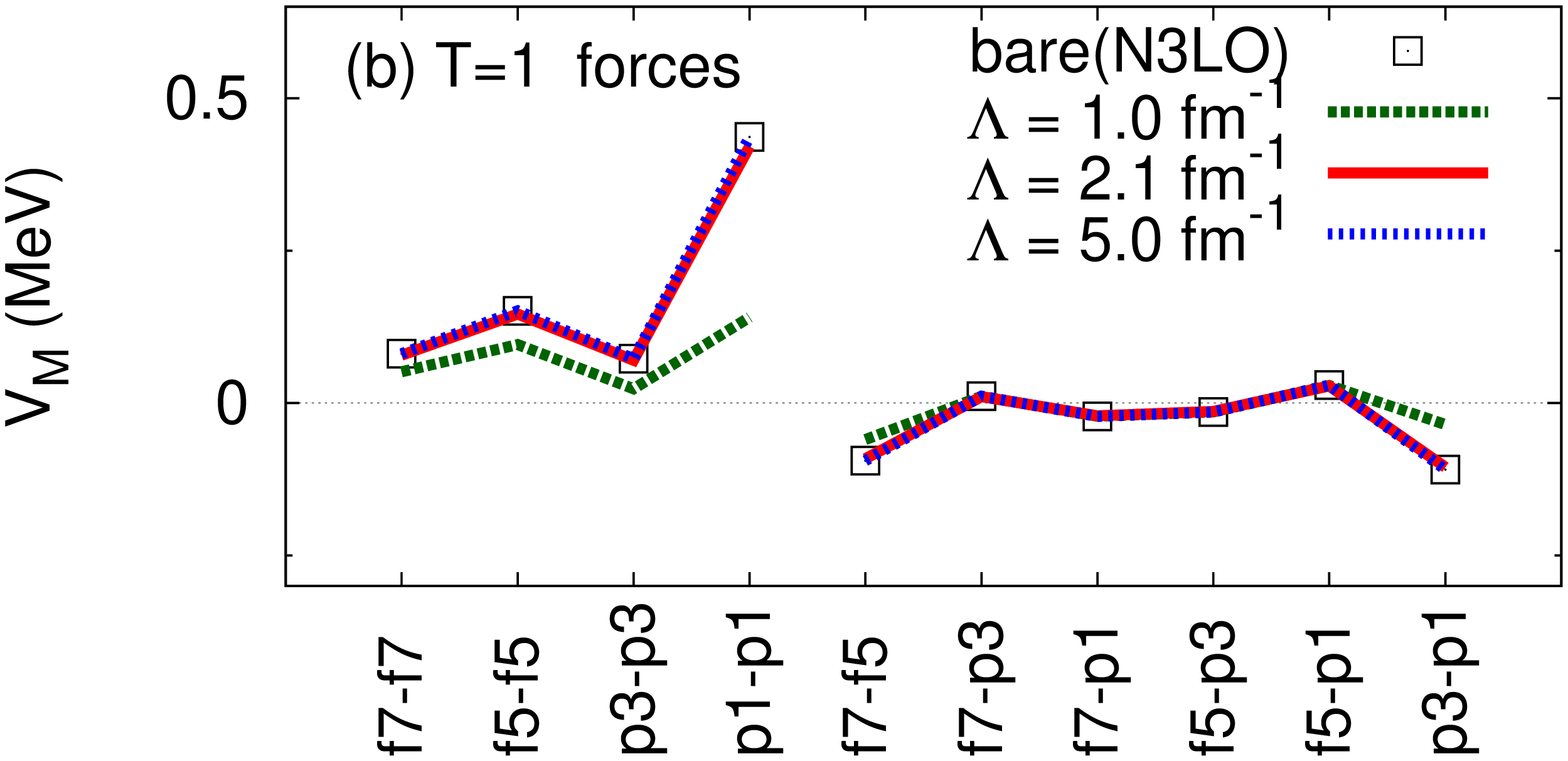}} & 
    \resizebox{72.25mm}{!}{\includegraphics[angle=270]
    {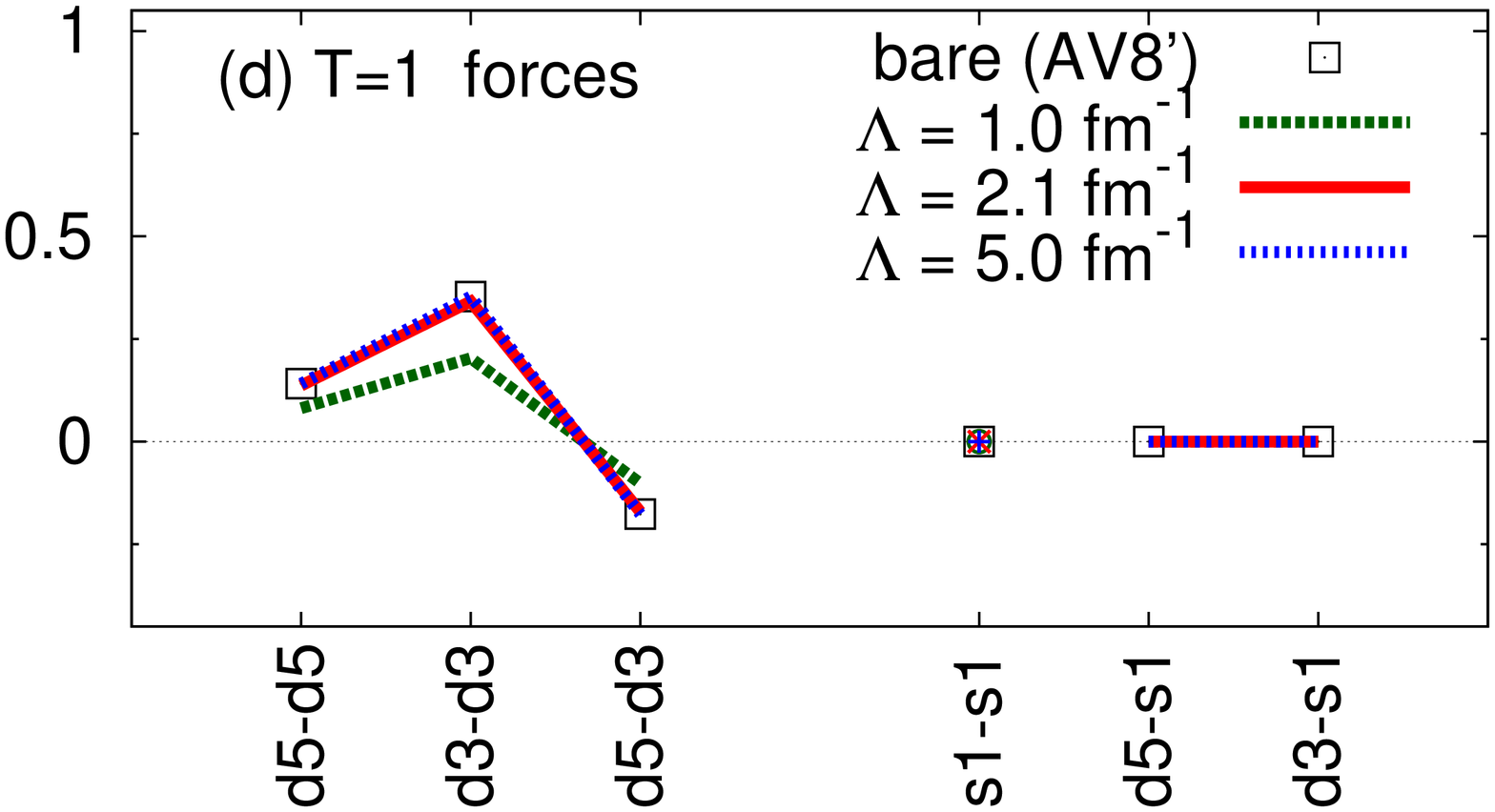}}\\ 
   \end{tabular}
   \caption{(color online.)
   Tensor-force monopole of $\Vlowk$ starting from the $\chi$N$^3$LO interaction
   with the same notation as in Fig.~\ref{fig:vlowk_ten}.}
   \label{fig:vlowk_ten_n3lo}
  \end{center}
 \end{figure*}

\begin{figure*}[htb]
  \begin{center}
   \begin{tabular}{cc}
    \resizebox{80mm}{!}{\includegraphics[angle=270]
    {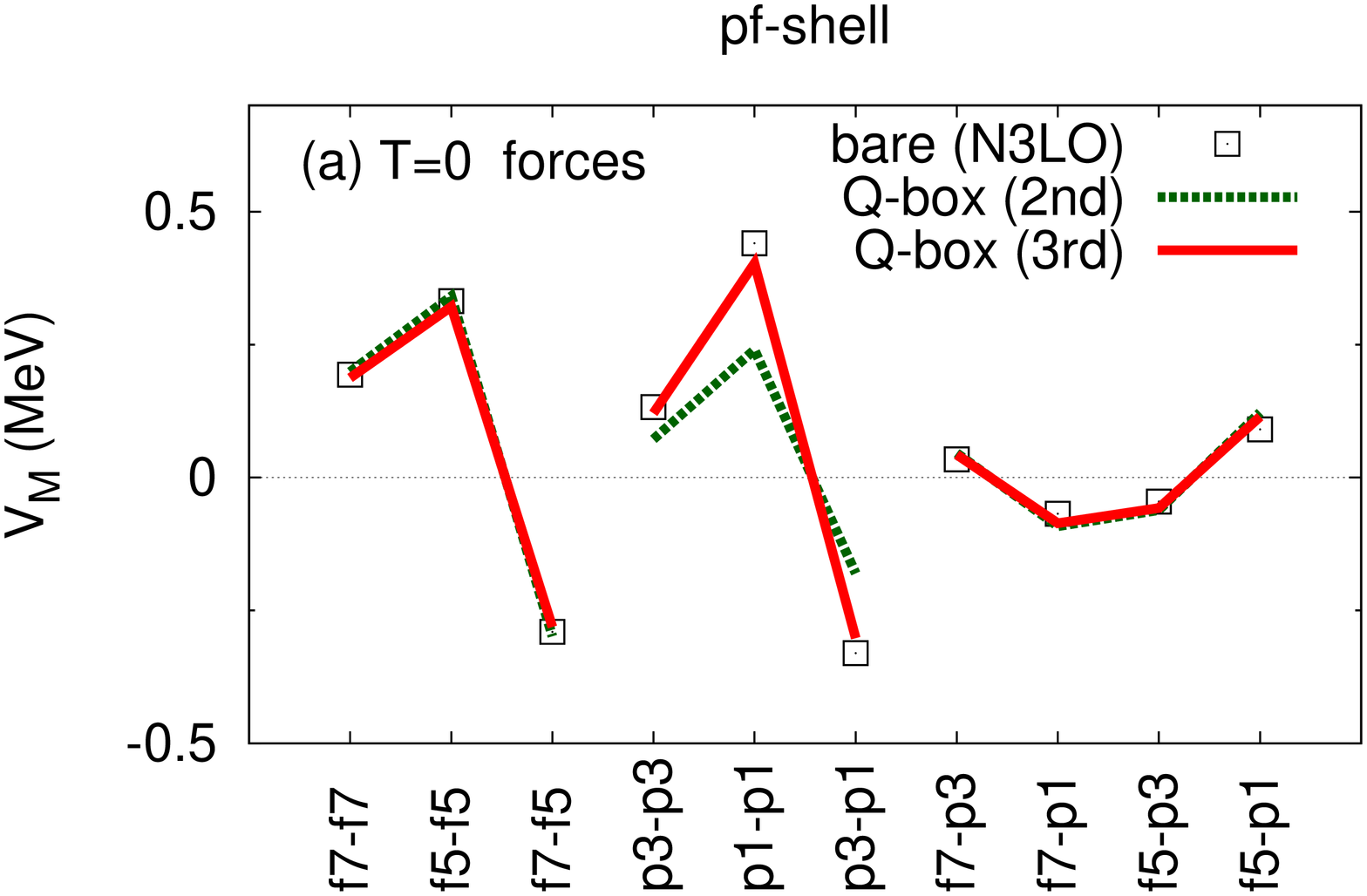}} & 
    \resizebox{72.25mm}{!}{\includegraphics[angle=270]
    {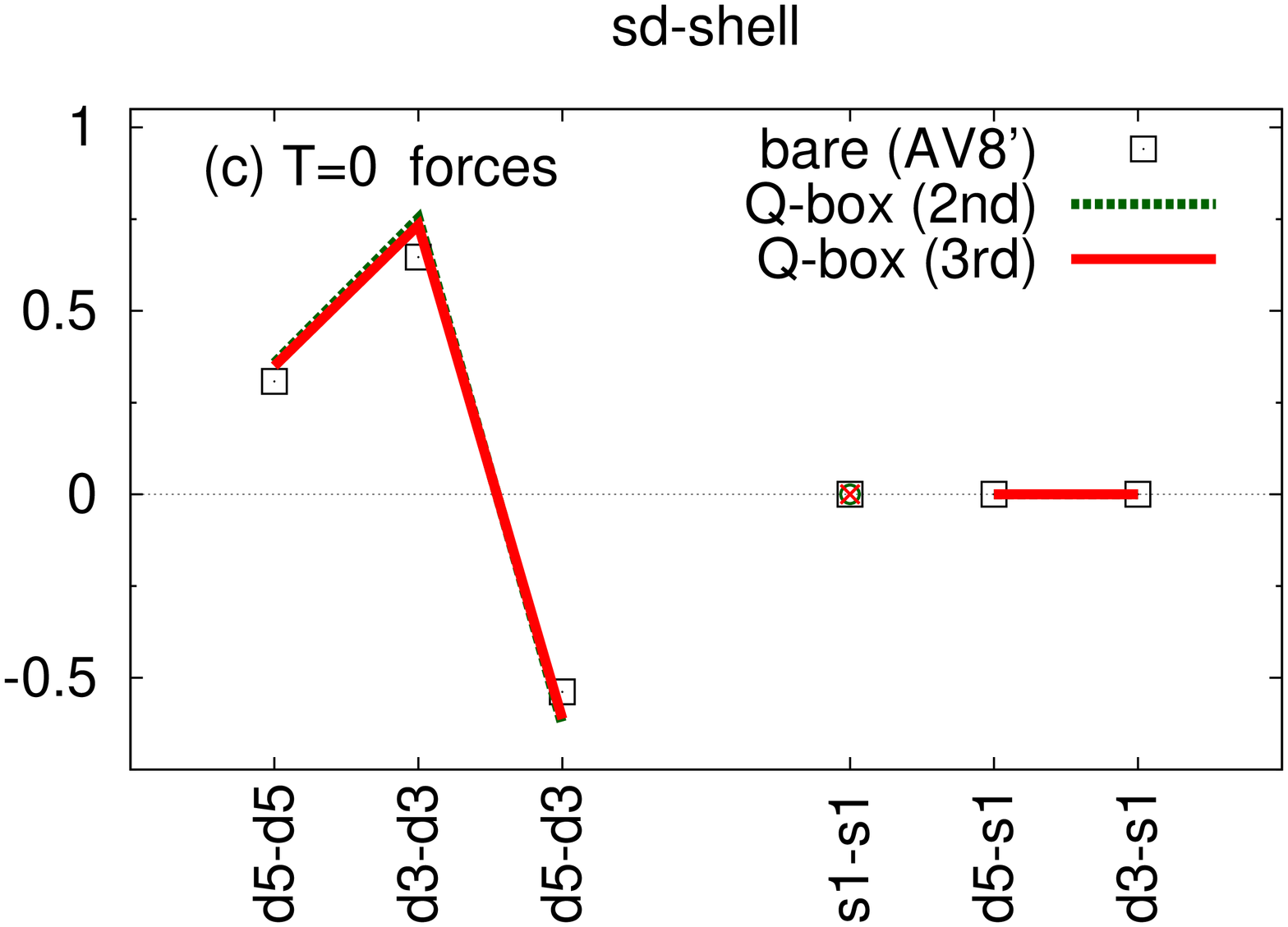}} \\
    \resizebox{80mm}{!}{\includegraphics[angle=270]
    {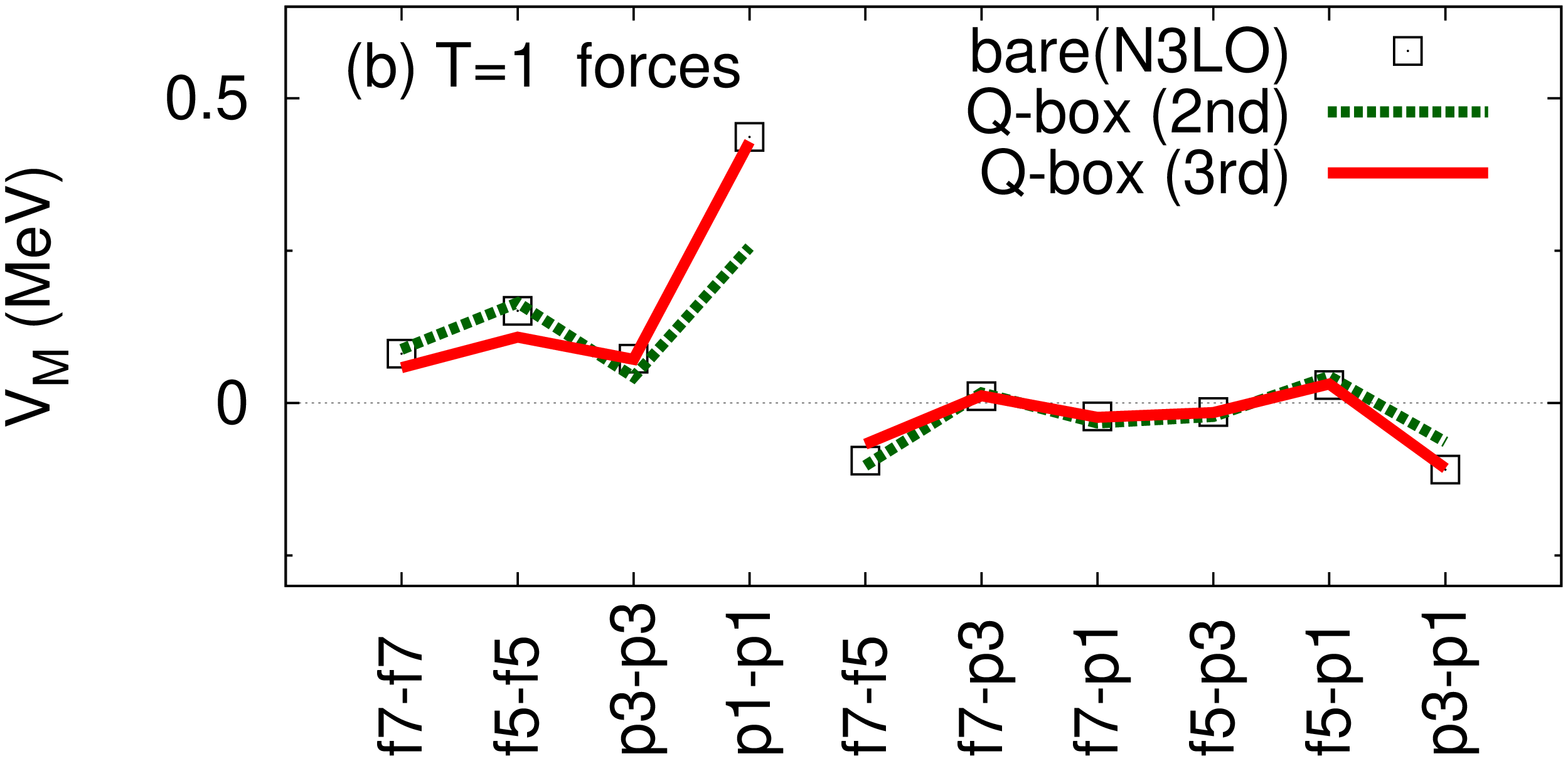}} & 
	\resizebox{72.25mm}{!}{\includegraphics[angle=270]
	{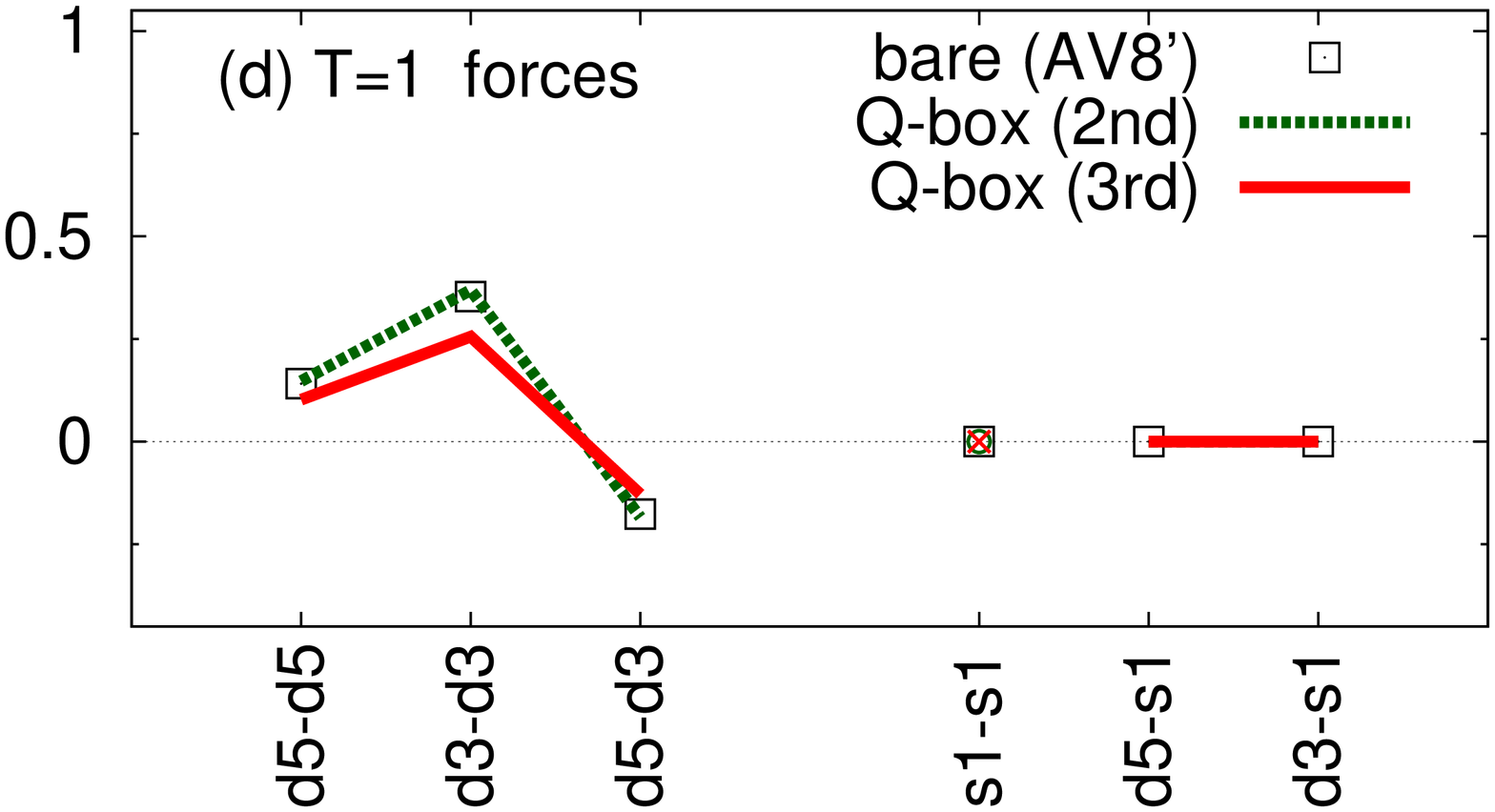}} \\
   \end{tabular}
   \caption{(color online.)
   Tensor-force monopole of $\Veffsm$ starting from the $\chi$N$^3$LO
   with the same notation as in Fig.~\ref{fig:Qbox_ten}.}
   \label{fig:Qbox_ten_n3lo}
  \end{center}
 \end{figure*}

\begin{figure*}[Htb]
  \begin{center}
   \begin{tabular}{cc}
    \resizebox{80mm}{!}{\includegraphics[angle=270]
    {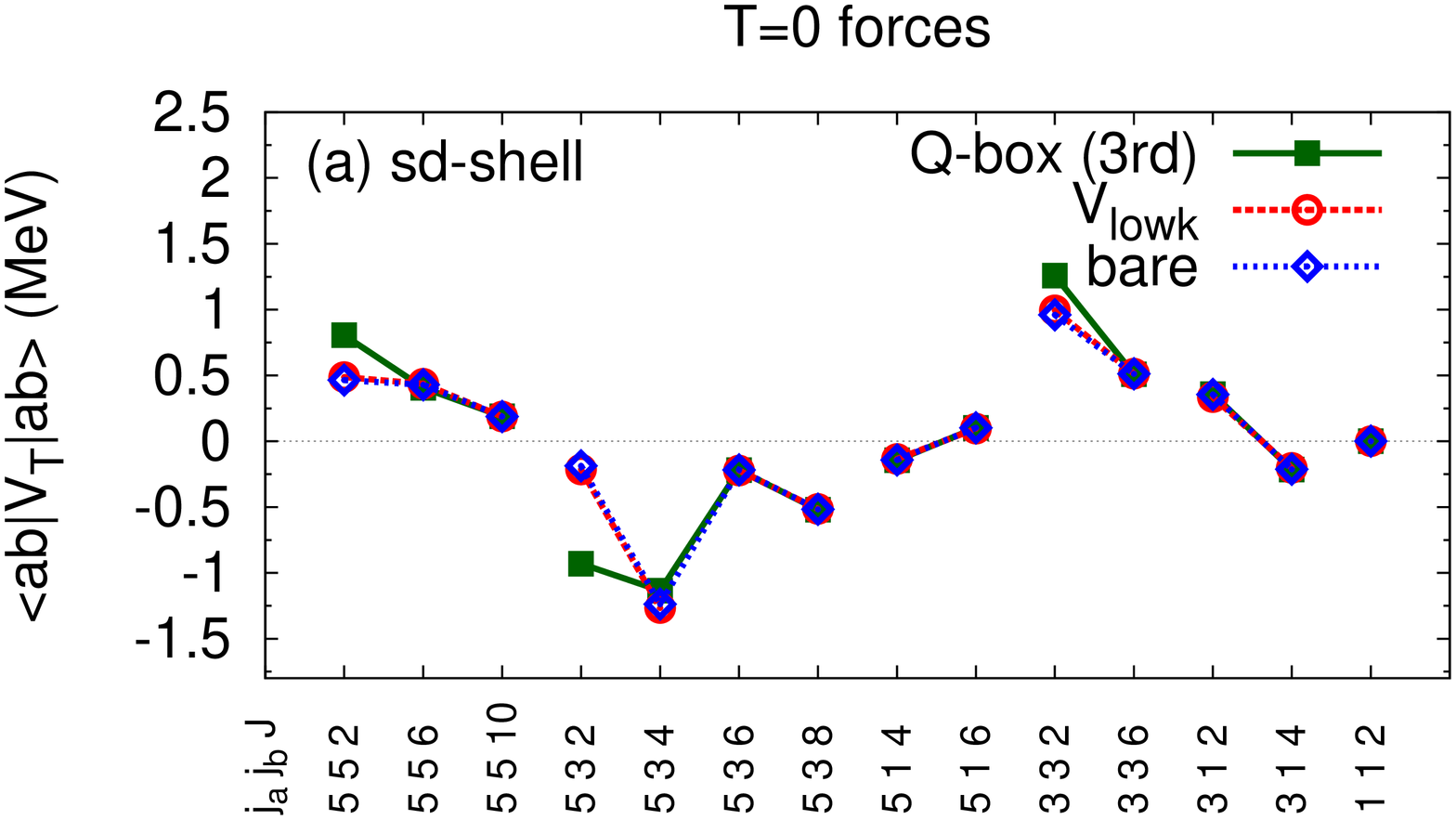}} & 
    \resizebox{72.25mm}{!}{\includegraphics[angle=270]
    {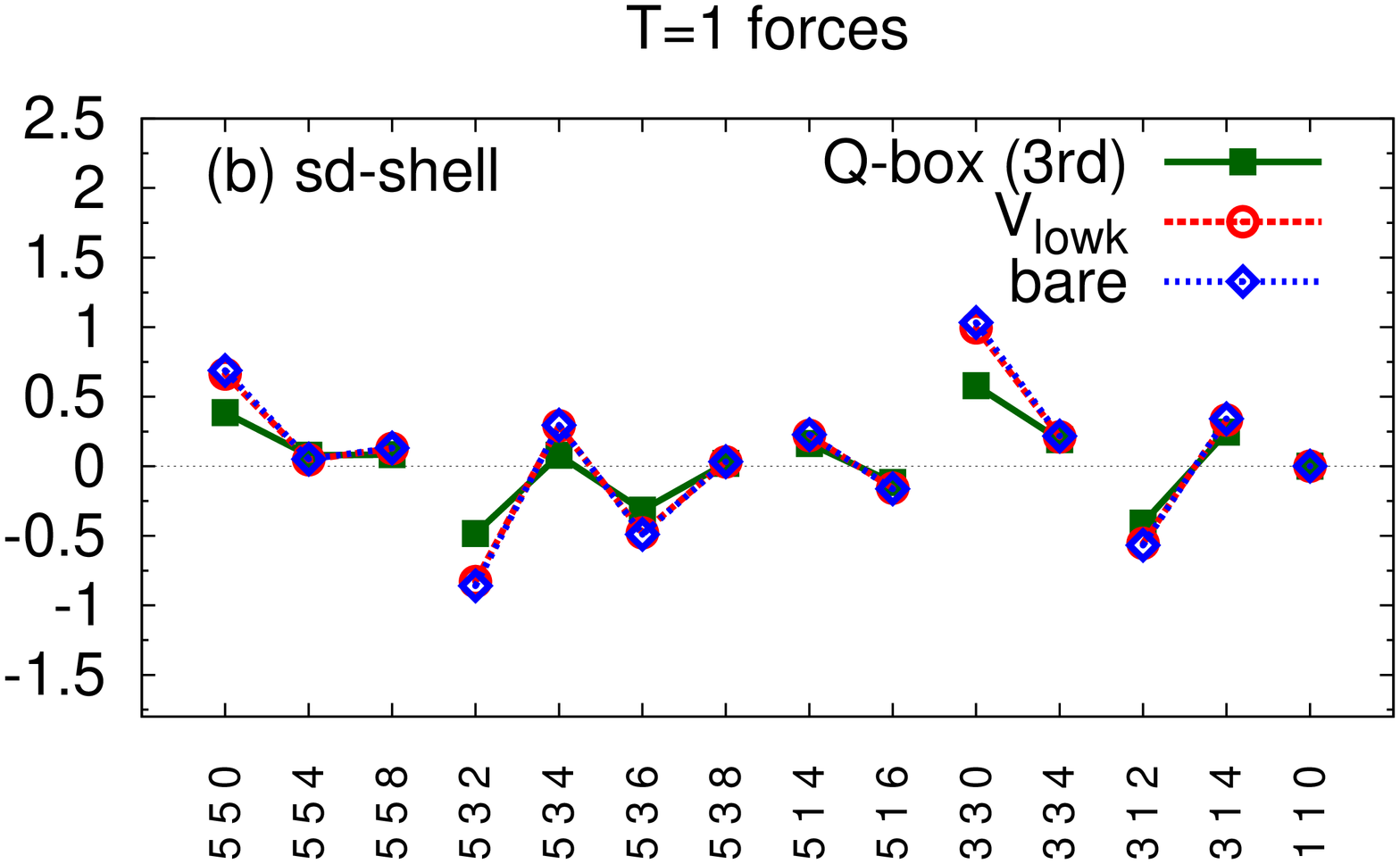}} \\ 
    \resizebox{80mm}{!}{\includegraphics[angle=270]
    {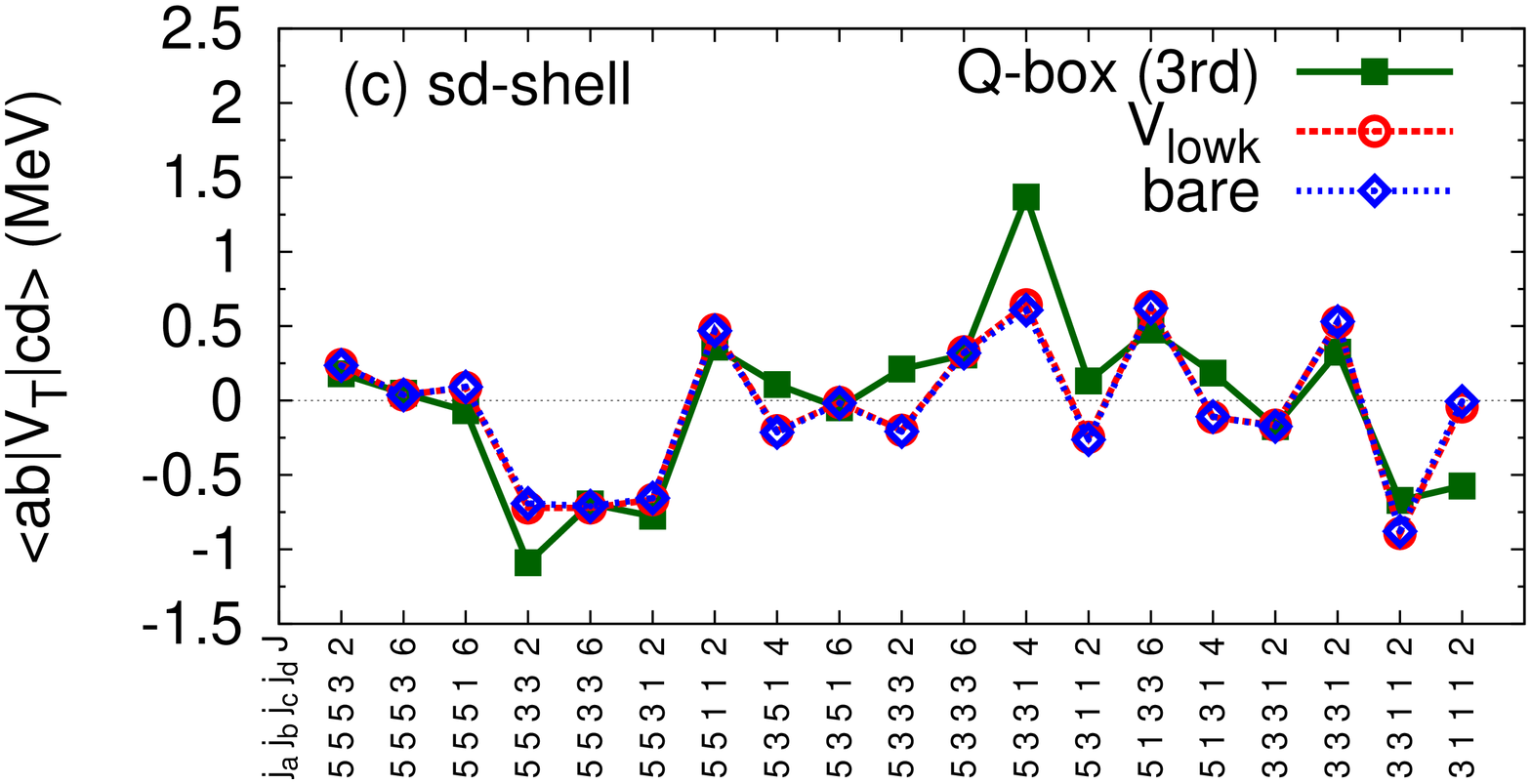}} & 
	    \resizebox{72.25mm}{!}{\includegraphics[angle=270]
	    {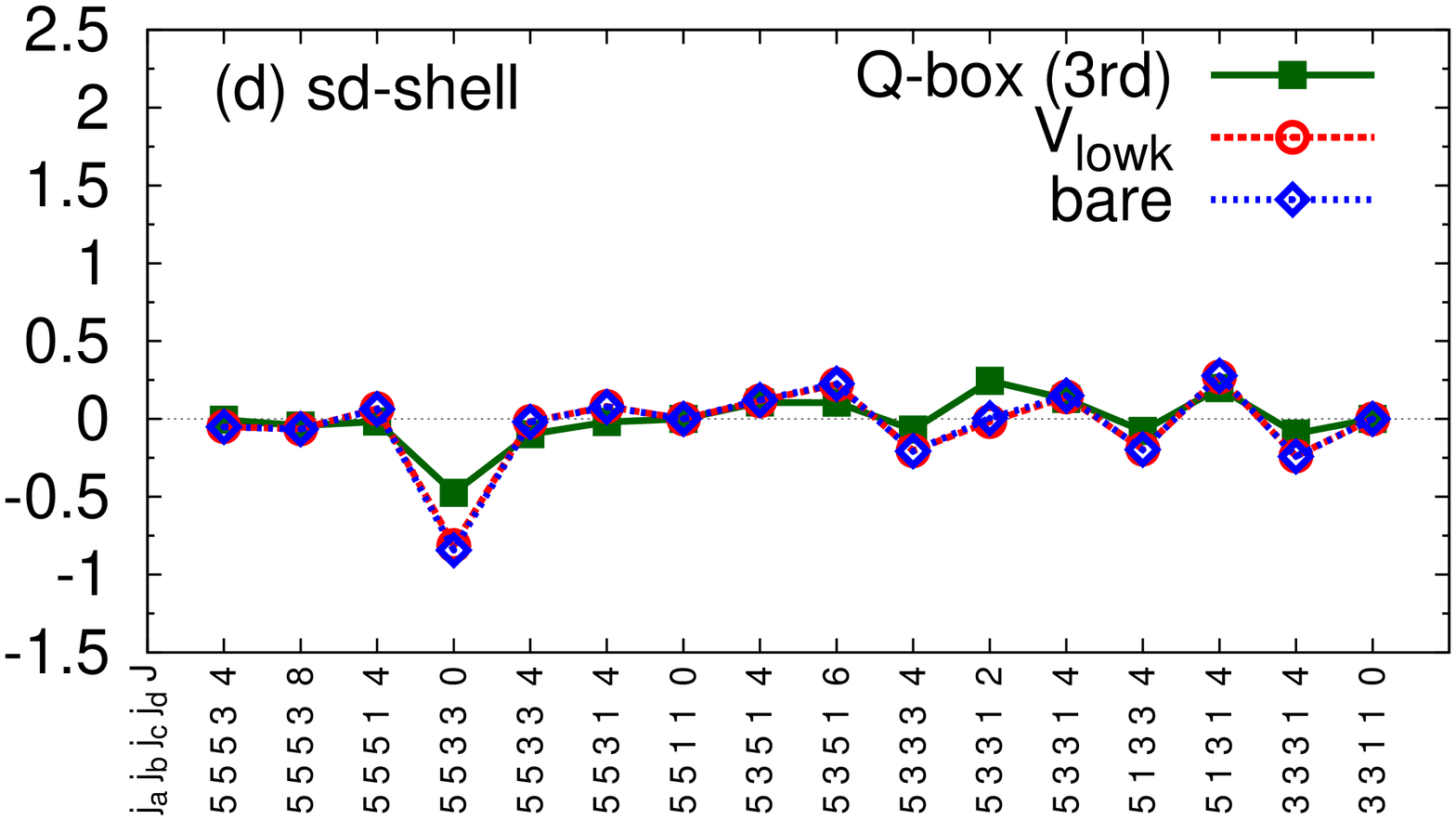}} \\ 
    \resizebox{80mm}{!}{\includegraphics[angle=270]
    {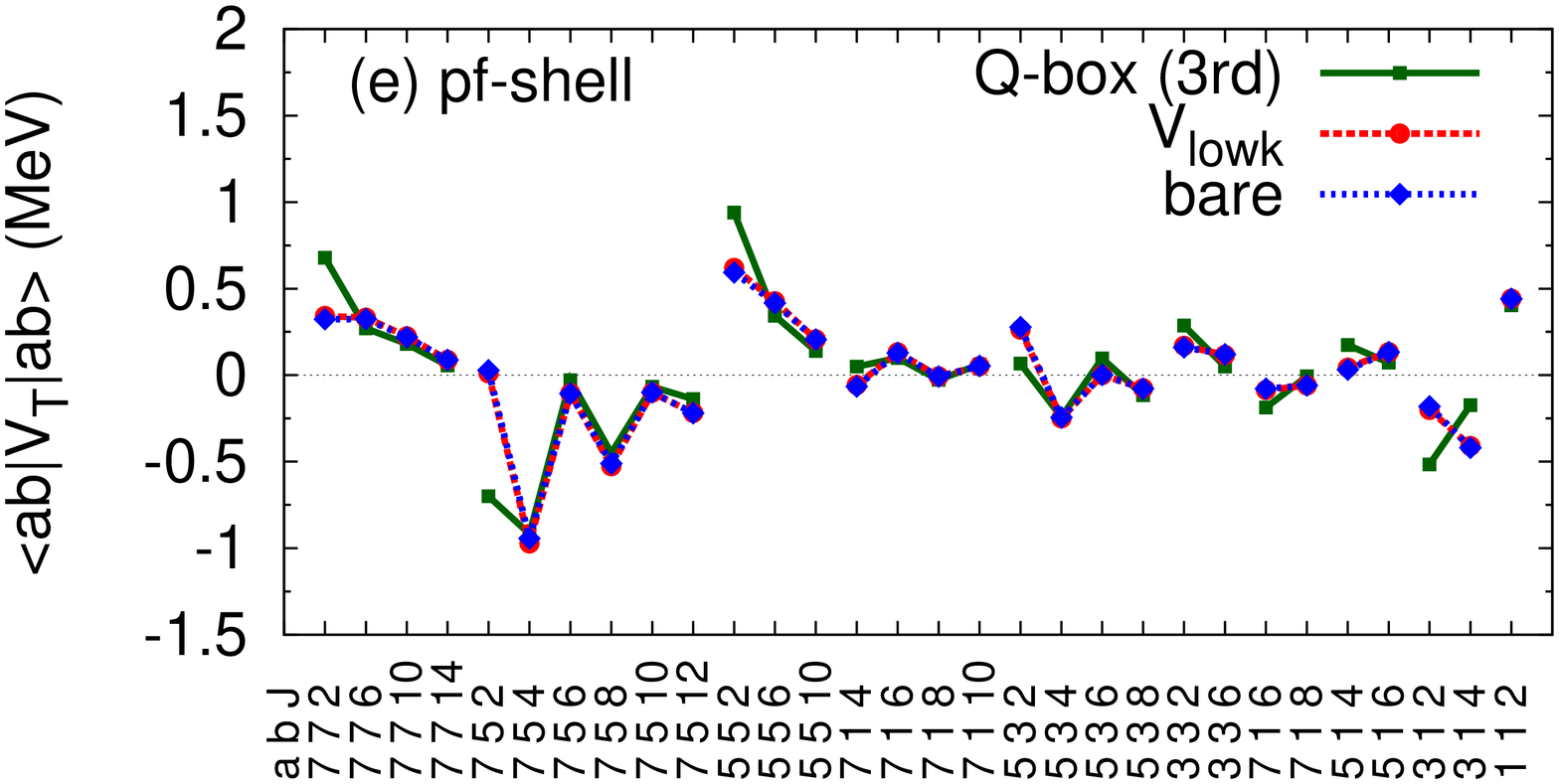}} & 
	\resizebox{72.25mm}{!}{\includegraphics[angle=270]
	    {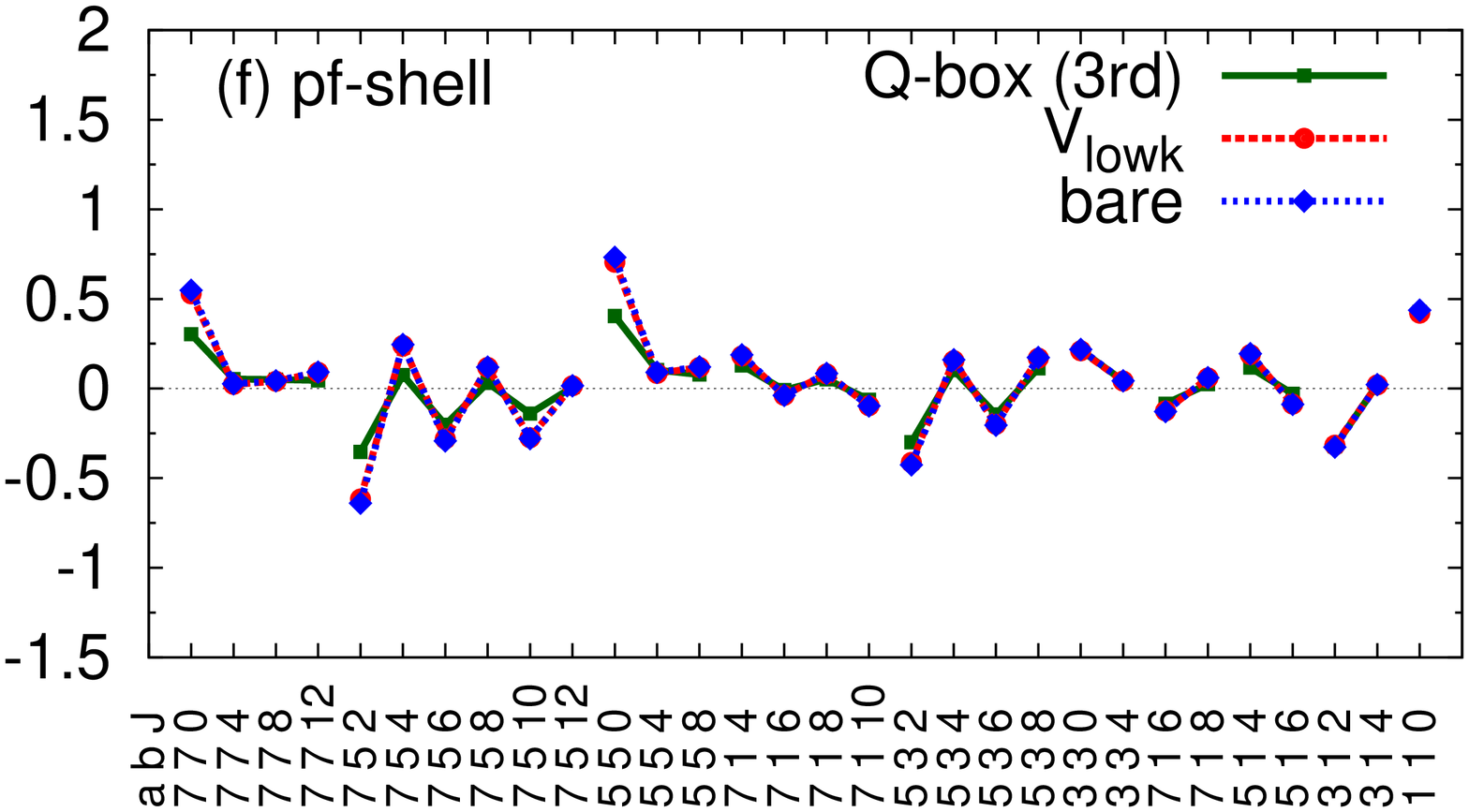}} \\ 
    \resizebox{80mm}{!}{\includegraphics[angle=270]
    {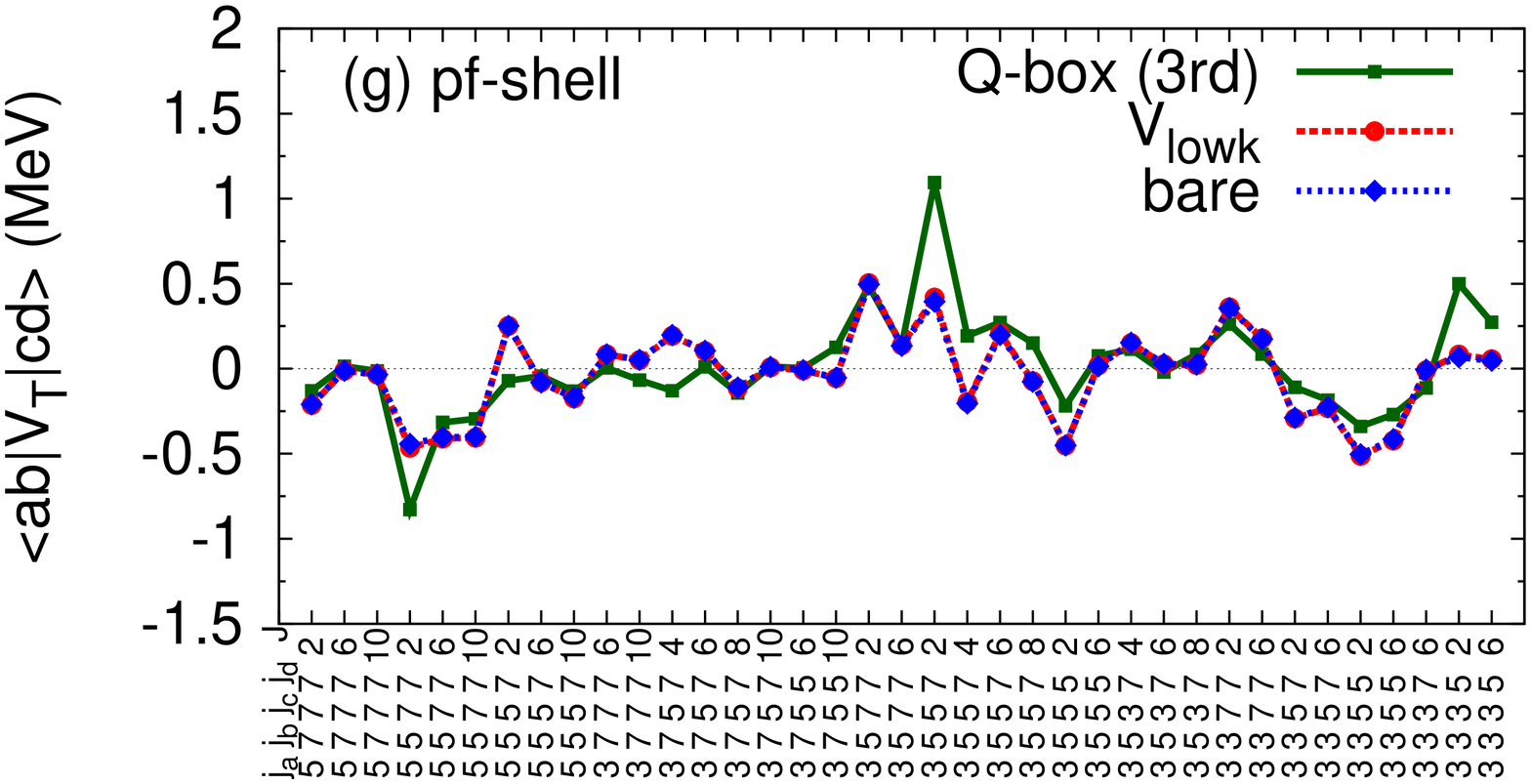}} &
	\resizebox{72.25mm}{!}{\includegraphics[angle=270]
	    {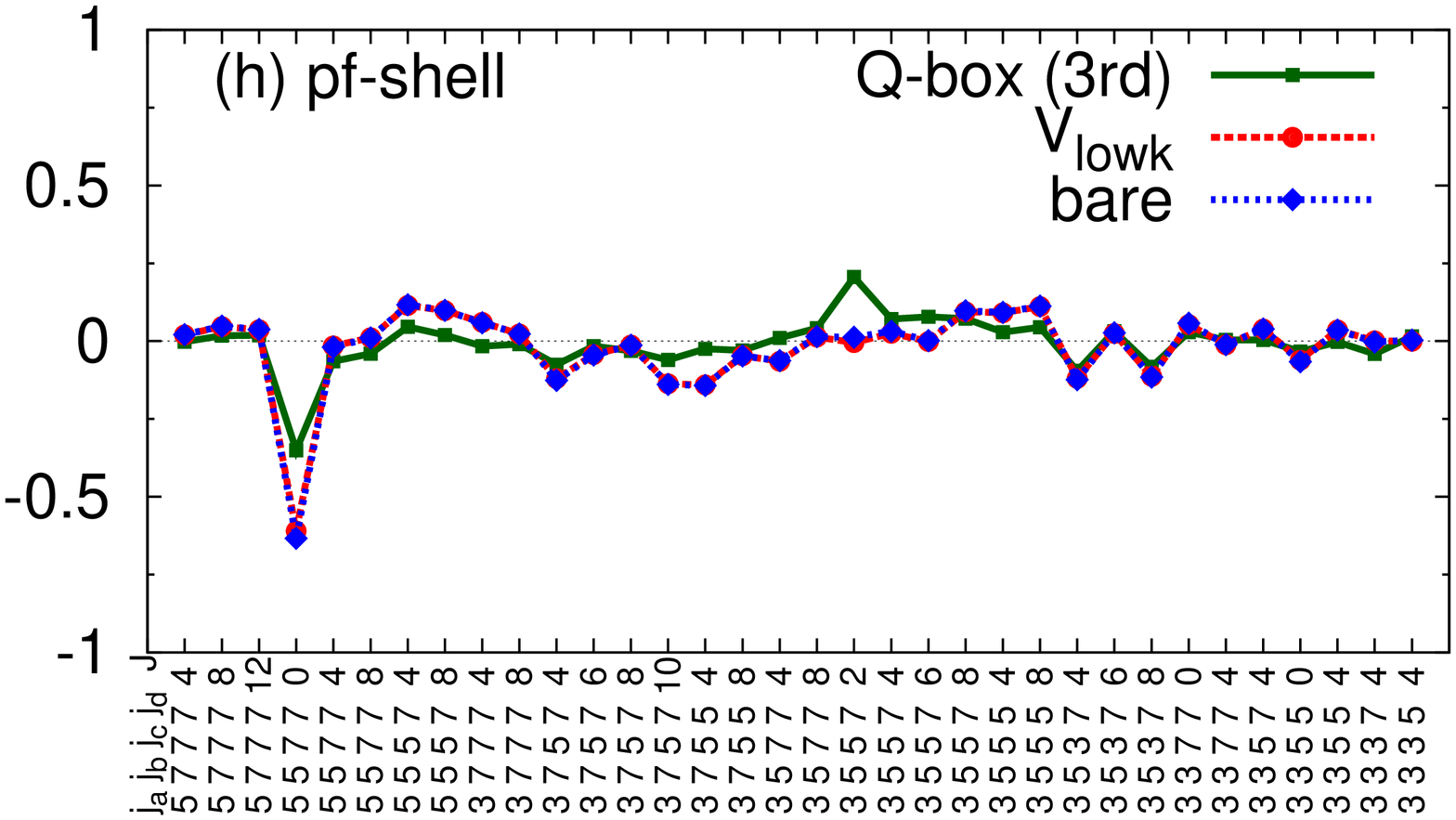}} \\ 
    \resizebox{80mm}{!}{\includegraphics[angle=270]
    {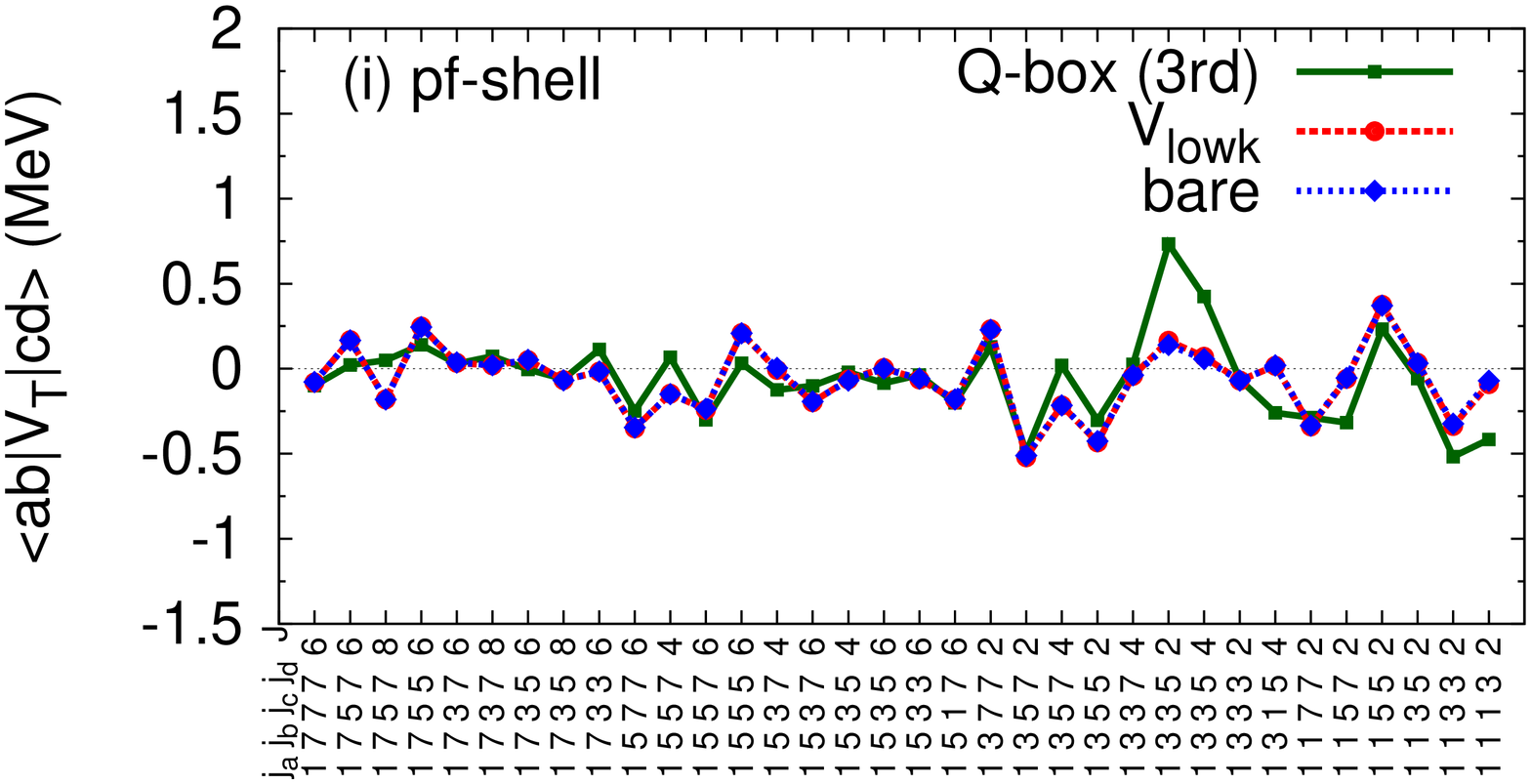}} & 
	\resizebox{72.25mm}{!}{\includegraphics[angle=270]
	{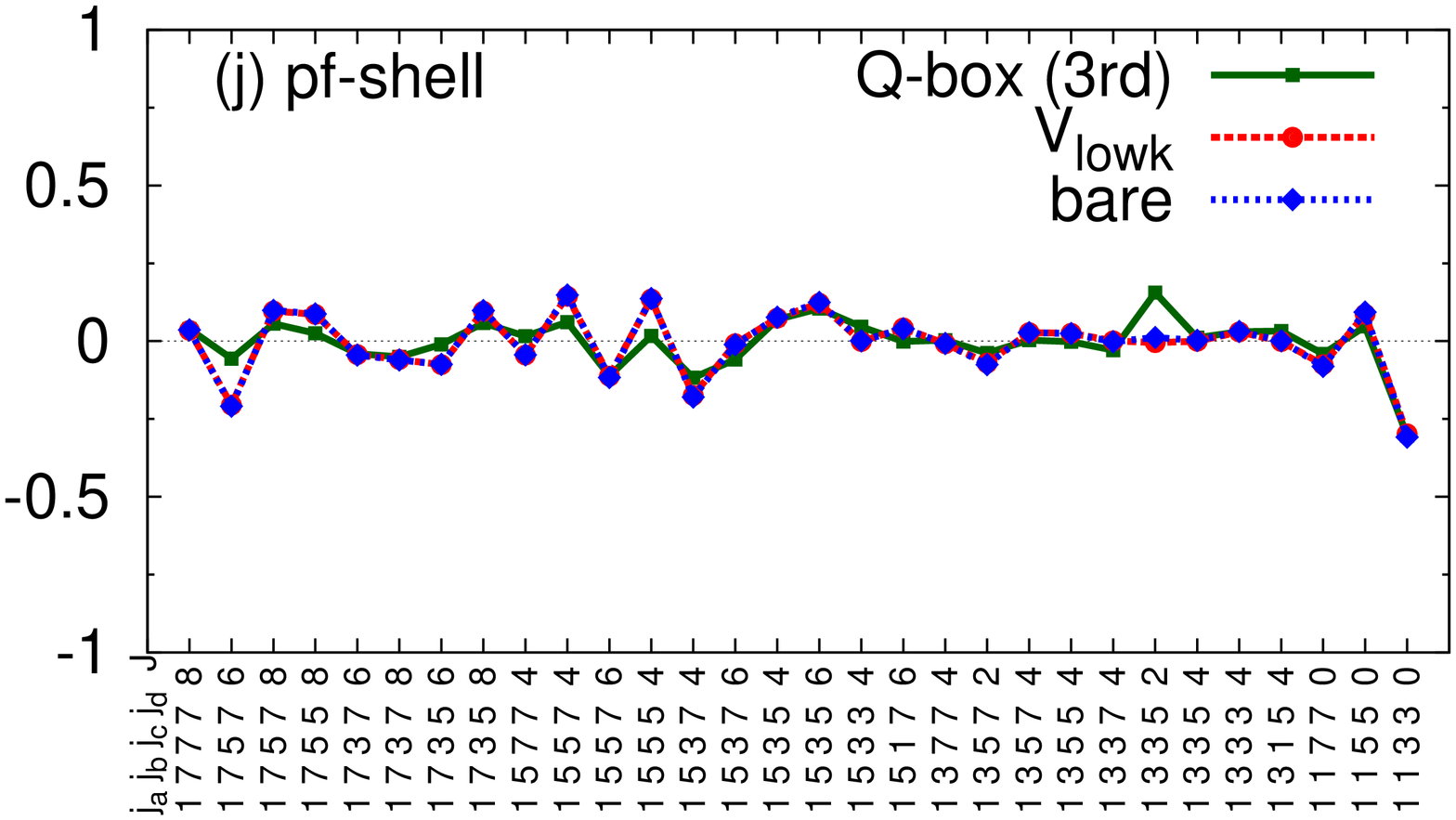}} \\ 
   \end{tabular}
   \caption{(color online.)
   Diagonal and non-diagonal matrix elements of the tensor-force component from
   effective interactions using the $\chi$N$^3$LO interaction. The labelling is the same as
   in Fig.~\ref{fig:multi_av8_d}.}
   \label{fig:multi_n3lo_od}
  \end{center}
 \end{figure*}

 We employ here another frequently used realistic interaction, $\chi$N$^3$LO,  
 as an example~\cite{RevModPhys.81.1773}.
 The $\chi$N$^3$LO interaction has a relatively smaller coupling between
 low-momentum and high-momentum modes compared with the AV8' potential.
 In Fig.~\ref{fig:vlowk_ten_n3lo} we show the monopole part of the
 tensor force of the $\chi$N$^3$LO bare potential and $\Vlowk$s with
 several cutoff parameters $\Lambda$($1.0~\fmi, 2.1~\fmi$ and
 $5.0~\fmi$). These results should be compared with the corresponding
 ones obtained with the AV8' interaction shown in
 Fig.~\ref{fig:vlowk_ten}. Figure~\ref{fig:Qbox_ten_n3lo} shows the
 tensor-force monopole of $\Veffsm$, corresponding to
 Fig.~\ref{fig:Qbox_ten} for the AV8' interaction, starting from the
 $\chi$N$^3$LO interaction. Finally, Fig.~\ref{fig:multi_n3lo_od} shows
 the multipole components of effective interactions, corresponding to
 Fig.~\ref{fig:multi_av8_d} for the AV8' interaction.
 
 In all the figures, we can conclude that all the
 features we discussed for the AV8' interaction pertain to the 
 $\chi$N$^3$LO interaction model as well. 
 
 \section{Conclusion}
In this work we have presented a detailed analysis of various contributions 
 to the nuclear tensor force as function of different renormalization
 procedures, starting with state-of-the-art nucleon-nucleon (NN)
 interactions and ending up with effective interactions for the nuclear 
 shell model. 
 The monopole part of the tensor force is weakly or barely 
 affected by various renormalization procedures, which in our case are 
 represented by a renormalization of the bare interaction and many-body
 perturbation theory in order to obtain an effective shell-model interaction. 
 This has lead us to introduce the concept of renormalization persistency 
 (R-Persistency) in the study of effective interactions.
 We studied the R-persistency of both renormalization procedures and showed 
 via numerical studies,
 their intuitive general explanations and
 a detailed algebraic analysis of core-polarization
 terms in perturbation theory,
 that this is a very robust process.
 We have also shown that the R-Persistency holds for
 two-body matrix elements including higher multipole components of the
 tensor force, although the deviation 
 increases somewhat if multipole components are included in the comparison.
 We conclude that the two renormalization steps (one for short-range 
 correlation and the other for in-medium effects) do not
 affect much either the monopole nor the multipole components
 of the tensor force, apart from slight differences between them. 
 Results obtained with two different interactions (AV8'
 and $\chi$N$^3$LO) lead us to the same conclusion, suggesting that 
 the R-Persistency of the tensor force for low-momentum states is a robust
 feature. This applies also to other interaction models than those studied here.

 The short-range part of the tensor force enters the
 renormalization of the central force, in particular in the $T=0$
 channel, producing on average an increased attraction. Since 
 the modification of the tensor force appears to be small, 
 the central force carries most of the renormalization effects 
 beyond first order in perturbation theory.

 Because the R-Persistency of the tensor force in effective 
 interactions is a robust feature, it may give a simple and concrete 
 starting point  for examining and constructing effective interactions, 
 especially phenomenological ones~\cite{Otsuka:2009qs}. In particular, 
 since the tensor force plays a significant role in the shell evolution 
 for nuclear systems with either large neutron/proton or proton/neutron 
 ratios, one can extract simple physics messages from complicated
 many-body systems.

 \begin{acknowledgments}
  We are very grateful to Professors R.~Okamoto and H.~Feldmeier
  for valuable discussions.
  This work is supported in part by Grant-in-Aid for Scientific
  Research (A) 20244022 and also by Grant-in-Aid for 
  JSPS Fellows (No.~228635), and by the JSPS Core to Core 
  program ``International Research Network for Exotic Femto Systems''
  (EFES).
 \end{acknowledgments}  
 
 

\end{document}